\documentclass{article}
\usepackage{amsmath,graphics,amssymb,amsfonts,latexsym,color,mathptmx,dsfont,bm}
\usepackage{nicematrix}
\usepackage{dsfont}
\newcommand{\Exp}{\mathds{E}}
\newcommand{\Var}{\mathds{V}\mbox{ar}}
\newcommand{\Cov}{\mathds{C}\mbox{ov}}
\newcommand{\vect}[1]{\vec{#1}}
\renewcommand{\vect}[1]{\boldsymbol{\bm #1}}
\newcommand{\mat}[1]{\boldsymbol{\bm #1}}
\DeclareMathOperator{\Po}{Po}
\newcommand{\Prob}{\mathds{P}}
\newcommand{\phasty}{\texttt{phasty}}
\usepackage{graphicx}
\usepackage{bm}
\newtheorem{theorem}{Theorem}[section]

\newtheorem{definition}[theorem]{Definition}

\usepackage{natbib}
\usepackage{subfig}
\newcommand\eqdef{\mathrel{\overset{\makebox[0pt]{\mbox{\normalfont\tiny\sffamily def}}}{=}}}
\newcommand{\ih}{{\mathrm{i}}}
\newcommand{\e}{{\mathrm{e}}}
\newcommand{\dd}{{\mathrm{d}}}
\DeclareBoldMathCommand\balpha{\alpha} 
\DeclareBoldMathCommand\bxi{\xi}
\DeclareBoldMathCommand\ba{a}
\DeclareBoldMathCommand\bc{c}
\DeclareBoldMathCommand\be{e}
\DeclareBoldMathCommand\bbm{m}
\DeclareBoldMathCommand\bp{p}
\DeclareBoldMathCommand\br{r}
\DeclareBoldMathCommand\bv{v}
\DeclareBoldMathCommand\bz{z}
\DeclareBoldMathCommand\bA{A}
\DeclareBoldMathCommand\bB{B}
\DeclareBoldMathCommand\bC{C}
\DeclareBoldMathCommand\bI{I}
\DeclareBoldMathCommand\bM{M}
\DeclareBoldMathCommand\bR{R}
\DeclareBoldMathCommand\bS{S}
\DeclareBoldMathCommand\bT{T}
\DeclareBoldMathCommand\bU{U}
\DeclareBoldMathCommand\bV{V}
\DeclareBoldMathCommand\bX{X}
\DeclareBoldMathCommand\bY{Y}
\DeclareBoldMathCommand\bZ{Z}
\DeclareBoldMathCommand\bzero{0}
\DeclareBoldMathCommand\bone{e}
\DeclareBoldMathCommand\bSigma{\Sigma}
\DeclareBoldMathCommand\bLambda{\Lambda}
\DeclareBoldMathCommand\bmu{\mu}
\DeclareBoldMathCommand\bDelta{\Delta}
\DeclareBoldMathCommand\bxi{\xi}
\renewcommand\ast{\prime}
\DeclareMathOperator{\diag}{diag}

\usepackage{bbm}
\newcommand{\indi}[1]{{\mathbbm{1}\left(#1\right)}}

\begin{document}
\title{Multivariate phase-type theory for the site frequency spectrum}
\date{\today}
\author{Asger Hobolth$^{1}$, Mogens Bladt$^{2}$ and Lars N{\o}rvang Andersen$^{3}$ \\
\small
1. Department of Mathematics, Aarhus University. 
Email: asger@math.au.dk \\
\small
2. Department of Mathematical Sciences, University of Copenhagen. 
Email: bladt@math.ku.dk \\
\small
3. Department of Mathematics, Aarhus University. 
Email: larsa@math.au.dk}
\maketitle
\section*{Abstract}
Linear functions of the site frequency spectrum (SFS) play a major role for understanding and investigating genetic diversity. Estimators of the mutation rate (e.g. based on the total number of segregating sites or average of the pairwise differences) and tests for neutrality (e.g. Tajima's~$D$) are perhaps the most well--known examples.
The distribution of linear functions of the SFS is important for constructing confidence intervals for the estimators, and to determine significance thresholds for neutrality tests. 
These distributions are often approximated using simulation procedures. 
In this paper we use multivariate phase--type theory to specify, characterize and calculate the distribution of linear functions of the site frequency spectrum.
In particular, we show that many of the classical estimators of the mutation rate are distributed according to a discrete phase--type distribution. 
Neutrality tests, however, are generally not discrete phase--type distributed. 
For neutrality tests we derive the probability generating function using continuous multivariate phase--type theory, and numerically invert the function to obtain the distribution. 
A main result is an analytically tractable formula for the probability generating function of the SFS. 
Software implementation of the phase--type methodology is available in the R package \phasty, and R code for the reproduction of our results is available as an accompanying vignette. \vspace{3mm} \\
{\bf Key words:}
Coalescent theory; mutation rate; phase--type distribution; site frequency spectrum.\\
{\bf Mathematical Subject Classification 2020:} Primary: 60J90. Secondary: 60J27, 60J28, 60J95, 92D15. 
\section{Introduction and motivation}
The Site Frequency Spectrum (SFS) $\bxi=(\xi_1,\ldots,\xi_{n-1})$ is a key quantity in population genetics. Entry $\xi_i$ is the number of sites in the locus under consideration where the mutant allele is present in $i$ of the $n$ samples; see 
e.g. Chapter~3~\&~4 in~\cite{wakeley2008coalescent}, Chapter~1~\&~2 in~\cite{Durrett2008} or Section 2.9 in~\cite{Etheridge2012}. The interest in $\bxi$ is because it is an easy summary statistics of an alignment of homologous DNA sequences. Several unbiased estimators of the mutation rate are given as linear combinations of the site frequencies \citep{achaz2009}, and similarly several tests of neutrality are also a weighted version of the SFS (e.g. \cite[\S 4.3]{wakeley2008coalescent}). In this paper we treat the multivariate distribution of $\bxi$ and the univariate distribution of weighted versions $\bc^\ast \bxi$, $\bc \in \mathbb{R}^{n-1}$, using phase-type theory. 
Phase-type distributions constitute a highly tractable class of probability distributions, and recently \cite{HSB2019} showed that the class arises naturally in coalescent theory. For example, the time to the most recent common ancestor and the total branch length are continuous phase--type distributed, and an entry in the SFS is discrete phase--type distributed.

We begin the paper with motivating examples that illustrate why univariate distributions of weighted versions of the SFS are of major interest: well-known estimators of the mutation rate $\theta$, and classical tests of neutrality, are of this form. Following \cite{Fu1994} we also determine the SFS-based unbiased estimator of the mutation rate with minimum variance. In the end of this introductory and motivating section we provide an overview of the remaining part of the paper. 
\subsection{Classical estimators of $\theta$ based on the SFS}
In the standard coalescent--with--mutation model  a number of unbiased estimators are suggested for the scaled mutation rate $\theta=4N\mu$. Here $2N$ is the number of chromosomes and $\mu$ is the mutation rate per generation for the locus under consideration. The estimators are based on the fact that the total branch length $Y_i$ in the coalescent tree with $i$ descendants has mean $2/i$ (e.g. \cite[\S 4.1.2]{wakeley2008coalescent}), and that mutations are sprinkled on the branches according to a Poission process with rate $\theta/2$. These two properties result in the expected SFS 
\begin{eqnarray}
  \label{meanSFS}
  \Exp[\xi_i]=
  \Exp\big[ \Exp[\xi_i|Y_i] \big]=
  \Exp\Big[ \frac{\theta}{2}Y_i \Big]=
  \frac{\theta}{2}\frac{2}{i}=
  \frac{\theta}{i},\;\;
  i=1,\ldots,n-1.
\end{eqnarray}
As noted in \cite[\S 2.2]{Durrett2008}, equation~(\ref{meanSFS}) implies that if we define a $(n-1)$-dimensional vector~$\bv$ with entries $v_i=1/i, \; i=1,\ldots,n-1,$ then for any 
$\bc = (c_1,\dots,c_{n-1}) \in \mathbb{R}^{n-1}$ with 
$\bc^\ast \bv = 1$ the linear combination
\begin{equation}
\bc^\ast \bxi = \sum_{i=1}^{n-1} c_i \xi_i \label{eq:basicunbiasedestimator}
\end{equation}
is an unbiased estimator of $\theta$. \cite{Ferretti229} provide an overview of well-known estimators of the form~\eqref{eq:basicunbiasedestimator}. The most simple estimators are perhaps the number of singletons 
\begin{eqnarray}
  \hat{\theta}_{\xi_1}=\xi_1,
  \label{SingletonEstimator}
\end{eqnarray}
and the scaled total number of segregating sites (Watterson's estimator)
\begin{eqnarray}
  \hat{\theta}_{\rm W}=\frac{1}{a_1} \xi_{\rm total}, \;\; {\rm where} \;\;
  a_1=\sum_{i=1}^{n-1} \frac{1}{i} \;\; {\rm and} \;\;
  \xi_{\rm total}=\xi_1+\cdots+\xi_{n-1}.
  \label{WattersonEstimator}
\end{eqnarray}   
Note that these two estimators are (up to a scaling factor) 0--1 weighted versions of the SFS. The coefficients for the singleton estimator are $c_1=1$ and $c_i=0, \; i=2,\ldots,n-1$, and for Watterson's estimator $c_i=1, \; i=1,\ldots,n-1$. Another classical estimator is based on the number of pairwise differences and is given by
\begin{eqnarray}
  \hat{\theta}_{\pi}=\frac{2}{n(n-1)} \sum_{i=1}^{n-1} i(n-i)\xi_i.
  \label{PairwiseDifferenceEstimator}
\end{eqnarray}
The coefficients of the SFS are (up to the scaling factor $2/\{n(n-1)\}$ given by $c_i=i(n-i), \;i=1,\ldots,n-1$, and thus this estimator is  a non--negative integer--weighted version of the SFS. Two other examples of non--negative integer--weighted estimators are
\begin{eqnarray}
  \hat{\theta}_{\rm H}=\frac{2}{n(n-1)} \sum_{i=1}^{n-1} i^2 \xi_i,
  \;\; {\rm and} \;\;
  \hat{\theta}_{\rm L}=\frac{1}{n-1} \sum_{i=1}^{n-1} i\xi_i,
  \label{HLestimators}
\end{eqnarray}
as suggested by \cite{Fay1405} and \cite{Zeng1431}.
\subsection{A best linear unbiased estimator (BLUE) of $\theta$ based on the SFS}  \label{BLUEsec}
Now suppose we wish to find the best linear unbiased estimator (BLUE) of $\theta$, i.e. an estimator of the form \eqref{eq:basicunbiasedestimator} with minimal variance. 
This estimator is the solution to the optimization problem
\begin{equation}
\begin{aligned}
& \underset{\vect{c}}{\rm minimize}
& & \Var[\vect{c}^{\ast} \vect{\xi}] \\
& {\rm subject} \; {\rm to}
& & \vect{c}^{\ast} \vect{v}=1.
\end{aligned}
\label{eq:optprob}
\end{equation}
Letting $\mat{\Lambda} = \Var[\vect{\xi}]$ denote the covariance matrix of $\bxi$,
we have $\Var[\vect{c}^{\ast} \vect{\xi}] = \vect{c}^{\ast} \mat{\Lambda} \vect{c}$
and \eqref{eq:optprob} is recognized as the minimization of a quadratic function
subject to a linear equality constraint. Standard methods 
(see e.g. \cite{BoydVandenberghe2004}) yields the solution
\begin{eqnarray}
  \hat{\bc}=
  \hat{\bc}(\theta)=
  \frac{\bLambda^{-1}\bv}{\bv^{\ast}\;\bLambda^{-1}\bv}.
  \label{BLUEestimator}
\end{eqnarray} 
We note that $\hat{\vect{c}}$ depends on $\theta$ through $\mat{\Lambda}=\mat{\Lambda}(\theta)$. 
\cite{Fu1994} suggested the estimator (\ref{BLUEestimator}), and 
\cite{FU1995172} showed how to calculate $\mat{\Lambda}(\theta)$ (see also \cite[\S 2.1]{Durrett2008}). The derivations by \cite{FU1995172} for $\mat{\Lambda}(\theta)$ are rather lengthy and specific to the standard coalescent with mutation model. In Section~\ref{generalCoefficients} below we show how multivariate phase--type theory can be used to determine the covariance matrix in a straight--forward manner.  

\begin{figure}[h]
\centering
\includegraphics[scale=0.34]{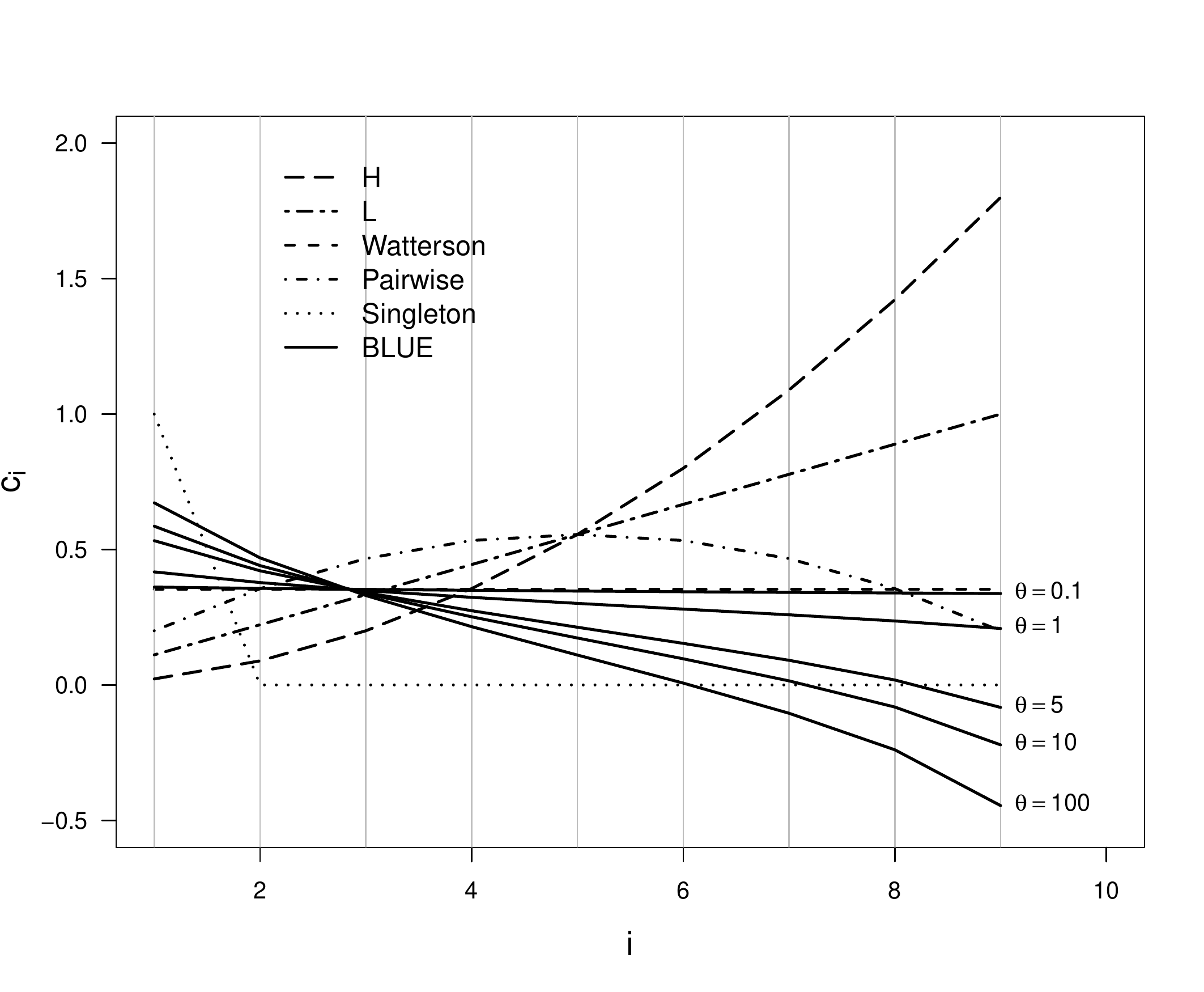}
\includegraphics[scale=0.34]{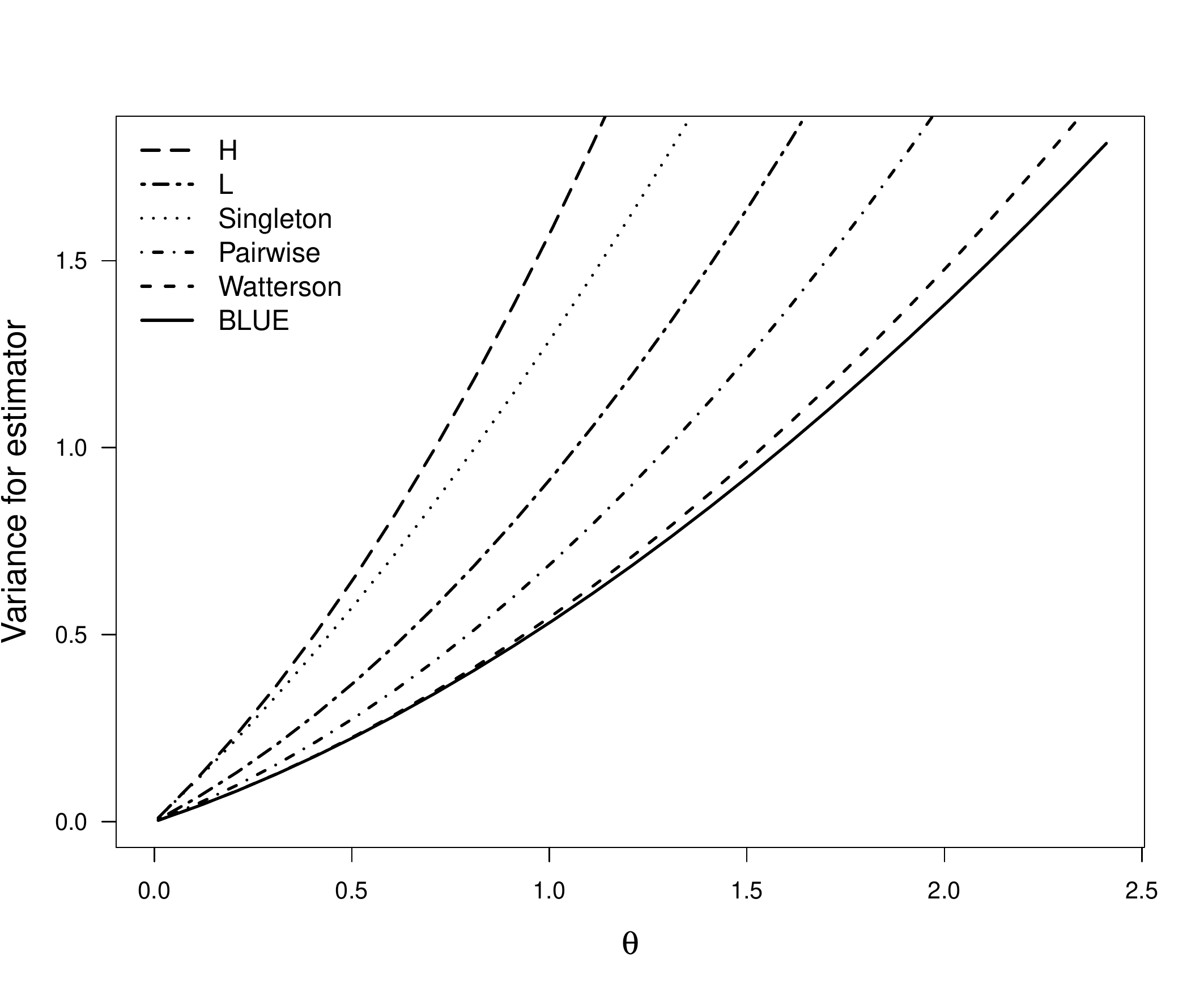}
\caption{Left: Coefficients $c_i$, $i=1,\ldots,n-1,$ for the SFS for various unbiased estimators (H, L, Wattersons, pairwise, singleton and BLUE) of $\theta$ for a sample of size $n=10$. Right: Variance $\Var[\vect{c}'\vect{\xi}]$ for the estimators as a function of $\theta$.}
\label{CoefficientsFigure}
\end{figure}

The left plot in Figure~\ref{CoefficientsFigure} shows the coefficients of the singleton estimator (\ref{SingletonEstimator}), Wattersons estimator (\ref{WattersonEstimator}), the pairwise difference estimator (\ref{PairwiseDifferenceEstimator}), the H- and L-estimators (\ref{HLestimators}), and the BLUE (\ref{BLUEestimator}) for a sample size of $n=10$. The coefficients for the BLUE are shown for various values of $\theta$. We note that the coefficients can be both positive and negative and are not necessarily integers. Furthermore we observe that Wattersons estimator coincides with the BLUE for small mutation rates (this result is shown formally in Section~\ref{generalCoefficients}), but for large mutation rates none of the classical estimators are similar to the BLUE.

The right plot in Figure~\ref{CoefficientsFigure} shows the variance for the estimators as a function of $\theta$. We note that the variance for Wattersons estimator is very similar to the  variance for small mutation rates, but for larger mutation rates (e.g. $\theta\geq 1$) the variance is noticeably smaller for the BLUE. We return to the BLUE for $\theta$ in Section~\ref{generalCoefficients}.

\subsection{Neutrality tests based on the SFS} \label{sec:neut}
Tests for neutrality are often based on the difference between two estimators of $\theta$ (e.g. \cite[\S 4.3.1]{wakeley2008coalescent}). For example, Tajima's~$D$ is essentially based on the difference between the pairwise difference estimator and Wattersons estimator $D=\hat{\theta}_{\pi}-\hat{\theta}_{\rm W}$. The neutrality test statistics are then also linear functions of the SFS, and access to the distribution allows us to provide significance values for rejecting the null model.          
\cite{Ferretti229} in their Table~3 provide an overview of neutrality tests based on unbiased linear estimators of $\theta$. In particular they consider $\hat{\theta}_{\pi}-\hat{\theta}_{\rm W}$, $\hat{\theta}_{\pi}-\hat{\theta}_{\rm H}$, $\hat{\theta}_{\rm L}-\hat{\theta}_{\rm W}$, $\hat{\theta}_{\rm W}-\hat{\theta}_{\rm H}$ and $\hat{\theta}_{\xi_1}-\hat{\theta}_{\rm W}$.

We note that the coefficient vector for the neutrality tests will contain both positive and negative entries. 
For instance for $n=4$ we get $a_1$ from \eqref{WattersonEstimator} to be $a_1=11/6$ and the coefficients for~$D$ are $c_i=i(4-i)/6-6/11$, $i=1,2,3$, or $c_1=c_3=1/2-6/11<0$ and $c_2=2/3-6/11>0$. 
For $n=8$ we get $a_1=140/363$ and the coefficients for $D$ are 
$c_i=i(8-i)/28-140/363$, $i=1,\ldots,7$, or  
$c_1=c_7=1/4-140/363<0$, $c_2=c_6=3/3-140/363>0$, $c_3=c_5=15/28-140/363>0$, 
and $c_4=4/7-140/363>0$.
In Section~\ref{sect:PoissonOnMPH} we provide the probability generating function (PGF) for a general weighted version of the SFS, and in~Section~\ref{generalCoefficients} we describe how to numerically invert the PGF. 

\begin{figure}[ptbh]
    \centering
    \subfloat[CDF for Tajima's $D$ with $n=4$ and $\theta=1$.]{{\includegraphics[width=7.5cm]{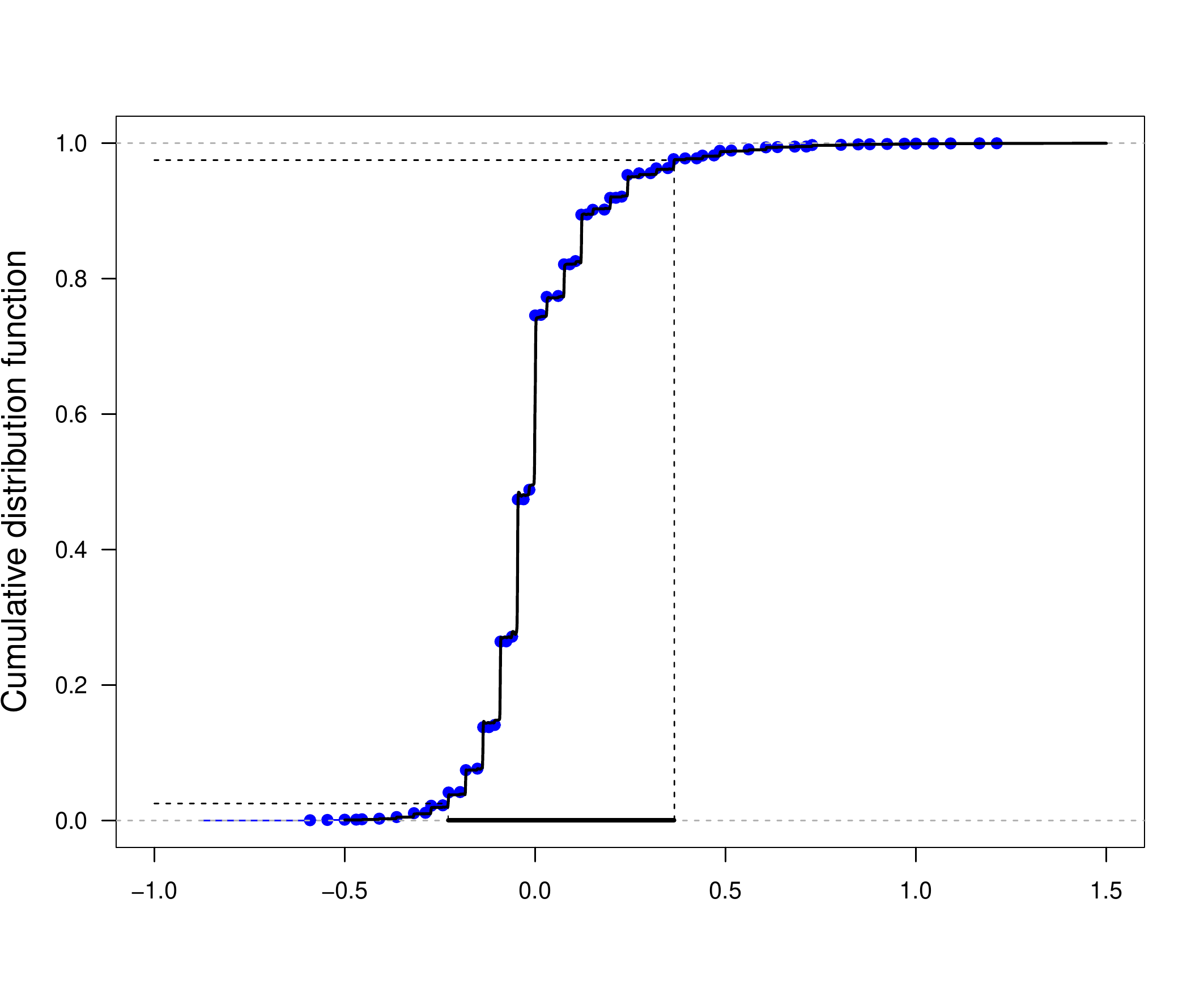} }}
    \quad
    \subfloat[CDF for Tajima's $D$ with $n=8$ and $\theta=1$.]{{\includegraphics[width=7.5cm]{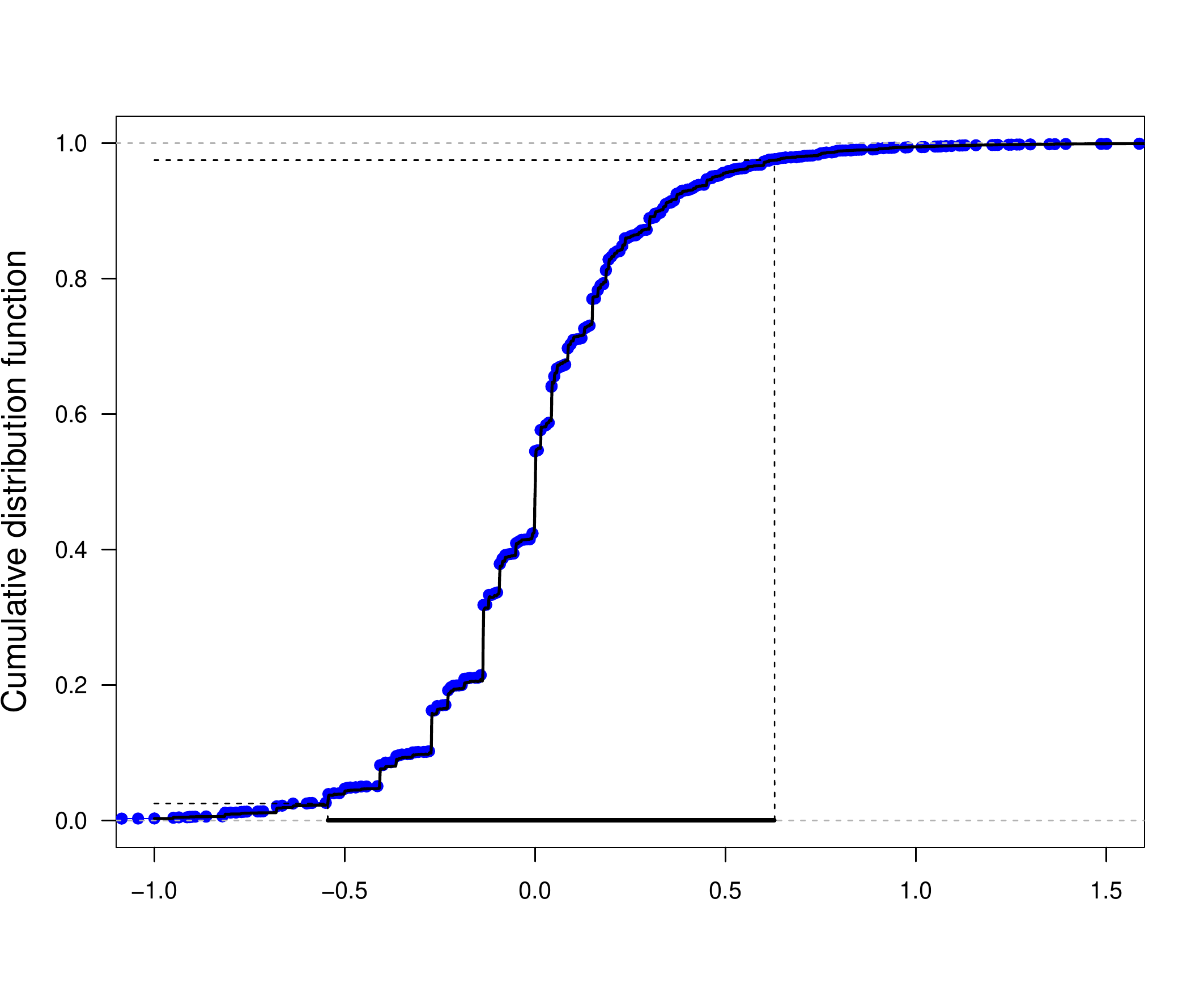} }}
    \caption{The black line is the approximation of the cumulative distribution function calculated using~(\ref{eq:ffteqn2}), the blue line is the empirical CDF of a sample of~$10.000$ simulated values of Tajima's $D$. We also emphasize the $2.5\%$ and $97.5\%$ quantiles.}
    \label{fig:cdfapprox}
\end{figure}

In Figure~\ref{fig:cdfapprox} we show the cumulative distribution function (CDF) of~$D$ for $n=4$ and $n=8$, and $\theta = 1$. 
The distribution is determined from a numerical inversion of the PGF.  
For both values of sample size~$n$, the result is compared to the empirical cumulative distribution function of a simulated sample. 

\subsection{Summary and organization of the paper}
In summary, estimators of $\theta$ can be divided into three classes depending on the type of coefficients of the SFS: (i) The coefficients are zero or one, $c_i \in \{0,1\}$; (ii) The coefficients are non--negative and integer--valued, $c_i \in \mathbb{N}_0$; or (iii) The coefficients are general, $c_i \in \mathbb{R}$. 
Classical estimators of $\theta$ are in class~(i) or~(ii), and the BLUE of $\theta$ falls in class~(iii).
Furthermore, neutrality tests also fall in class~(iii).

In this paper we characterize and calculate the distribution of a linear function of the SFS in each of these three cases. We apply and develop phase--type theory (e.g. \cite{bladt-nielsen-2017}; \cite{HSB2019}) for this purpose. 
A particular focus is on the distribution of a Poisson distributed stochastic variable with a phase--type distributed rate. 
The reason is that the multivariate phase-type distribution is a framework for the ancestral process (coalescent tree) of the samples, and mutations are sprinkled on the tree according to a Poisson process. 
A main result in this paper is a simple formula for the probability generating function (PGF) for the SFS. Actually, we believe this analytically tractable result is a major step forward for future mathematical treatments of the SFS for the standard coalescent and beyond.    

In Section~\ref{sect:PH}, we introduce phase--type theory with a view towards applications in population genetics. In the remainder part of the paper we capitalize on Section~\ref{sect:PH}. 
Firstly, we consider variables in class~(i) and show that they are discrete phase--type (DPH) distributed. 
In particular, in Section~\ref{01coefficients}, we introduce the \textit{block-counting process} and build upon Theorem 3.5 in \cite{HSB2019}. 
Secondly, we show that variables in class~(ii) are also DPH--distributed, but the construction is substantially more complicated than for class~(i).
The construction is described in Section~\ref{integerCoefficients}.
Thirdly, variables of class~(iii) are in general {\em not} DPH--distributed, but the probability generating function (PGF) is available in closed form. We describe how to calculate and invert the PGF in order to obtain the full distribution. Access to the full distribution for an estimator of $\theta$ allows us to calculate confidence intervals for the parameter, and access to the full distribution of a neutrality test allows us to specify significance thresholds.
In Section~\ref{generalCoefficients} we consider general coefficients, and apply the framework to determine the BLUE of the mutation rate and for neutrality tests. 

The paper ends with a discussion on extensions to more general coalescent models.
\section{Preliminaries: Phase--type theory} \label{sect:PH} 
In this section we establish notation and review some results from phase-type theory. 
First, we introduce the continuous phase-type (PH) and discrete phase-type (DPH) distributions. Both distributions are absorption times of Markov processes, and both are analytically tractable with closed expressions (via simple matrix manipulations) for the probability distribution functions as well as probability generating functions, Laplace transforms and moments. 
We refer to \cite{bladt-nielsen-2017} for more information about PH-distributions.
\subsection{The discrete phase--type (DPH) distribution: \\
  Definition, distribution and probability generating function}
Consider a time-homogeneous discrete-time Markov chain $\{X_i\}_{i\in \mathbb{N}}$ with state space $\{1,2,\ldots,p,p+1\}$, where the first~$p$ states are transient and the last state $p+1$ is absorbing. Write the transition probability in the form
\begin{eqnarray}
  \mat{P}=
  \begin{pmatrix}
    \mat{M} & \vect{m} \\
    \vect{0} & 1 \\
  \end{pmatrix},
\end{eqnarray} 
where $\mat{M}$ is the $p \times p$ matrix that determines the transition probabilities between the transient states, $\vect{0}$ is a row vector of zeros, and $\vect{m}=\vect{e}-\mat{M}\vect{e}=(\mat{I}-\mat{M})\vect{e}$ is the vector of transition probabilities to the absorbing state. Here~$\vect{e}$ is the column vector of ones. The vector of initial probabilities is $\vect{\pi}=(\pi_1,\ldots,\pi_p)$, such that $\mathbb{P}(X_0=i)=\pi_i$.

\begin{definition}[Discrete Phase-type distribution]
The time to to absorption 
\begin{equation}
  \tau_d=\min \{i\geq 0:X_i=p+1\}, \label{eq:dphdef}
\end{equation}
is said to have a discrete phase--type distribution of order~$p$ with initial distribution $\vect{\pi}$ and sub--transition probability matrix $\mat{M}$, 
and we write $\tau_d\sim\mbox{DPH}_p(\vect{\pi},\mat{M})$. 
\end{definition}
Note that with this definition we always have $\tau_d \geq 1$. The probability mass function of the DPH--distribution is 
$\mathbb{P}(\tau_d =0)=0$ and
\begin{equation}
  \Prob(\tau_d =i)=\vect{\pi}\mat{M}^{i-1}\vect{m}, \;\; i=1,2,\ldots. \label{eq:dphmpf}
\end{equation}
The probability generating function (PGF) is 
\begin{eqnarray}
  \Exp[z^{\tau_d}] = \sum_{i=1}^{\infty} z^i \vect{\pi} \mat{M}^{i-1}\vect{m}
  =z\vect{\pi} \Big( \sum_{i=0}^{\infty} (z\mat{M})^i \Big) \vect{m}  
  =z\vect{\pi} (\mat{I}-z\mat{M})^{-1} \vect{m}
  =z\vect{\pi} (\mat{I}-z\mat{M})^{-1}(\mat{I}-\mat{M})\vect{e},
  \;\; -1<z<1.
  \label{DPHpgf}
\end{eqnarray}
The factorial moments are found by differentiating the PGF with respect to~$z$ and evaluating in $z=1$. The moments are provided in Theorem~1.2.69 in~\cite{bladt-nielsen-2017}. In particular the mean is given by 
\begin{eqnarray}
  \Exp[\tau_d]=\vect{\pi}(\mat{I}-\mat{M})^{-1}\vect{e},
  \label{DPHmean}
\end{eqnarray} 
and the second factorial moment is
\begin{eqnarray}
  \Exp[\tau_d(\tau_d-1)]=2\vect{\pi}\mat{M}(\mat{I}-\mat{M})^{-2}\vect{e}.
  \label{DPHfact2}
\end{eqnarray} 
\subsection{The phase--type (PH) distribution: Definition, density and Laplace transform}
A phase--type distribution is the time to absorption of a Markov jump process. More formally, 
consider a continuous--time Markov jump process $\{ X_t\}_{t\geq 0}$ with finite state-space $\{ 1,2,...,p,p+1\}$, where states $1,...,p$ are transient and state $p+1$ is absorbing. This means that $\{ X_t\}_{t\geq 0}$ has an intensity matrix $\mat{\Lambda}$ of the form 
\begin{equation}
 \mat{\Lambda} =\begin{pmatrix}
\mat{S} & \vect{s} \\
\vect{0} & 0
\end{pmatrix} , \label{eq:lambda-matrix}
\end{equation}
and we refer to the $p\times p$ sub-matrix of rates between the transient states $\mat{S} =\{ s_{ij}\}_{i,j=1,...,p}$
as a {\it sub-intensity} matrix, the $p$-dimensional column vector $\vect{s}=(s_i)_{i=1,...,p}$  as an {\it exit rate} vector (since its elements are the intensities for jumping to the absorbing state), and finally $\vect{0}$ is a $p$-dimensional row vector of zeros. The assumption of states $1,...,p$ being transient means that eventually the process will jump to the absorbing state. Since rows sum to zero in intensity matrices (i.e. $\mat{\Lambda}\vect{e}=\vect{0}$), row sums are non--positive (zero or negative) in sub--intensity matrices. Furthermore, from $\mat{\Lambda}\vect{e}=\vect{0}$ we get $\vect{s}=-\mat{S}\vect{e}$. Hence the exit rate vector $\vect{s}$ is easy to determine from the sub-intensity matrix $\mat{S}$. 

Assume that $\{ X_t\}$ begins in a transient state and let $\vect{\alpha}=(\alpha_1,...,\alpha_p)$ where  $\alpha_i=\Prob (X_0=i)$, $i=1,...,p$. Then $\vect{\alpha}\vect{e}=\sum_{i=1}^p \alpha_i=1$ and $\vect{\alpha}$ is a probability vector on the set of transient states $\{ 1,2,\dotsc,p\}$. Often $\vect{\alpha}=\vect{e}_1=(1,0,...,0)$, i.e. the process begins in state~1. 

\begin{definition}[Phase-type distribution]
The time until absorption
 \[  \tau = \inf\{ t>0 : X_t = p+1  \} \]
is said to have a phase-type distribution of order $p$ with initial distribution $\vect{\alpha}$ and sub-intensity matrix $\mat{S}$, and we write 
\[ \tau\sim \mbox{PH}_p(\vect{\alpha},\mat{S}) . \]
\end{definition}

The probability for the Markov jump process to be in the different transient states is determined by a matrix exponential. We have 
\begin{equation}
 \Big( \Prob (X_t=1),..., \Prob (X_t=p) \Big)  = \vect{\alpha}e^{\mat{S}t}. 
 \label{eq:defective-dist-at-time-t}
 \end{equation}
We say that $\vect{\alpha}e^{\mat{S}t}$ is the {\it defective} distribution of $X_t$ on $\{1,2,...,p\}$ since the probabilities do not sum to one due to the possibility of having been absorbed prior to time $t$.
We see that
\[  \Prob (\tau>t) = \sum_{i=1}^p \Prob (X_t = i) =\sum_{i=1}^p (\vect{\alpha}e^{\mat{S}t})_i =  \vect{\alpha}e^{\mat{S}t}\vect{e}. \]
Hence the distribution function for $\tau$ is
\[ F(t)=1-\Prob (\tau>t) = 1- \vect{\alpha}e^{\mat{S}t}\vect{e}, \]
and we get the density 
\begin{eqnarray*}
  f(t) = \frac{d}{dt} F(t) = -\vect{\alpha}e^{\mat{S}u}\vect{S}\vect{e}
  = \vect{\alpha}e^{\mat{S}u} \vect{s} . 
\end{eqnarray*}  
The expected time spent in state~$j$ given the initial state is~$i$ is given by
\begin{eqnarray*}
  U_{ij}=\Exp\Big[\int_0^{\tau} \indi{X_t=j}dt|X_0=i\Big]
        =\int_0^{\infty}\Prob(X_t=j|X_0=i)dt
        =\int_0^{\infty}(e^{\mat{S}t})_{ij}dt
        =\Big( [\mat{S}^{-1}e^{\mat{S}t}]_0^{\infty} \Big)_{ij}
        =(-\mat{S}^{-1})_{ij}. 
\end{eqnarray*}
Here, $\mat{S}$ is invertible because all eigenvalues for sub--intensity matrices have strictly negative real parts (e.g. Corollary 3.1.14 in \cite{bladt-nielsen-2017}).
The matrix $\mat{U}=(-\mat{S})^{-1}$ is called the Green matrix.
The Laplace transform for $\tau$ is given by
\begin{eqnarray}
  \mathcal{L}_\tau (t)
  =\Exp[ e^{-\tau t}]
  =\int_0^\infty e^{-tx}\vect{\alpha}e^{\mat{S}x}\vect{s}dx 
  =\vect{\alpha}\left( \int_0^{\infty} e^{-(t\mat{I}-\mat{S})x} dx \right)\vect{s}
  =\vect{\alpha}(t\mat{I}-\mat{S})^{-1}\vect{s},
  \;\; t \geq 0.
  \label{LaplacePH}
\end{eqnarray}
Here, $t\mat{I}-\mat{S}$ is invertible because the real part of all eigenvalues are strictly larger than~$t$. 
From the Laplace transform we obtain the moments $\mu_n$ of $\tau$ by differentiating and evaluating in zero
\begin{equation}\label{moments}
  \mu_n = \Exp [\tau^n] = (-1)^n L^{(n)}_{\tau}(0) = 
  (-1)^n (-1)^n n! \vect{\alpha} (-\mat{S})^{-n} \vect{e} =  
  n! \vect{\alpha}(-\mat{S})^{-n} \vect{e}=
  n! \vect{\alpha} \bU^n \vect{e},
\end{equation}
where $\bU=(-\bS)^{-1}$ is the Green matrix.
Note for future reference that 
\begin{eqnarray}
  \mu_1=\Exp[\tau]=\balpha \bU \be
\end{eqnarray}
and 
\begin{eqnarray}
  \mu_2=\Exp[\tau^2]=2\vect{\alpha}\mat{U}^2 \vect{e}.
\end{eqnarray}  
\subsection{Transformations using rewards} \label{sec:rewards}
Let $\tau \sim \mbox{PH}_p(\vect{\alpha},\mat{S})$ and $\{ X_t\}_{t\geq 0}$ its underlying Markov jump process. We define a reward function 
\begin{eqnarray}
  r: \{ 1,\ldots,p \} \rightarrow \mathbb{R}_{+}, 
\end{eqnarray}
and let $\br = (r(1),\ldots,r(p))^\ast=(r_1,\dots,r_p)^\ast \in \mathds{R}_+^p$ be the vector of non-negative rewards. 
We then define the total reward $Y$ earned before time $\tau$ as
\begin{equation}
  Y=\int_0^\tau r(X_t)dt.  \label{eq:PH-reward}
\end{equation}
Letting $\br = \bone$ we recover $\tau$, so the class of distributions defined by~\eqref{eq:PH-reward} contains the PH-distributions. 
A rather remarkable fact is that when we restrict
ourselves to non-negative rewards, we will remain within the class of PH-distributions. In fact, for positive rewards ($r_i > 0$ for all $i$), 
it is straight--forward to show (e.g. \cite{HSB2019}) that 
\begin{equation}
  Y \sim \mbox{PH}_p(\vect{\alpha},\mat{\Delta}(\br)^{-1}\mat{S})\, ,
  \label{PosRewards}
\end{equation}
where $\bDelta(\br) \eqdef \diag(r_1,\dots,r_p)$ is the diagonal matrix whose non-zero entries are given by $\br$.
For non-negative rewards where some rewards are zero, the construction is more involved. As described in \cite[\S 3.1.8]{bladt-nielsen-2017} the random variable $Y$ of (\ref{eq:PH-reward}) is then a mixture distribution of a point mass at 0 and a phase-type distribution.

Consider the embedded Markov chain $\{ Y_n\}_{n\in\mathds{N}}$ with transition matrix $\mat{Q}=\{q_{ij}\}_{i,j=1,\ldots,p}$ where $q_{ij}=-s_{ij}/s_{ii}$ for $i\neq j$, and $q_{ii}=0$.  
Define $E^+=\{ i\in E: r_i>0 \}$ and $E^0 = \{ i\in E: r_i=0\}$ and decompose accordingly the vector $\vect{\alpha}=(\vect{\alpha}^+,\vect{\alpha}^0)$ and  
\[ \mat{Q} = \begin{pmatrix}
\mat{Q}^{++} & \mat{Q}^{+0} \\
\mat{Q}^{0+} & \mat{Q}^{00}
\end{pmatrix} . \]
Let $d=|E^+|$ be the number of elements in $E^+$ and define
\begin{eqnarray}
  \mat{P}=\mat{Q}^{++}+\mat{Q}^{+0}
  \left( \mat{I} - \mat{Q}^{00}\right)^{-1}\mat{Q}^{0+} \;\; {\rm and} \;\;
  \vect{\pi} = \vect{\alpha}^+ + \vect{\alpha}^0
  (\mat{I}-\mat{Q}^{00})^{-1}\mat{Q}^{0+}.
  \label{ZeroConstruction}
\end{eqnarray}
Then $\mat{P}=\{ p_{ij}\}_{i,j=1,...,d}$ is the transition matrix of the Markov chain which is obtained from $\{ Y_n\}_{n\in \mathds{N}}$ at times when $Y_n\in E^+$. This follows by noticing that the $(i,j)$'th element of
$\mat{Q}^{+0}(\mat{Q}^{00})^n\mat{Q}^{0+}$ is the probability of going from $i$ to $j$ by first making a transition to a state in $E^0$, remaining in $E^0$ for the next $n$ jumps, and finally jumping from a state in $E^0$ to $j$, and since
\[ \left( \mat{I} - \mat{Q}^{00}\right)^{-1} = \sum_{m=0}^{\infty} (\mat{Q}_{00})^m . \]
With a similar argument, $\pi_i$ gives the probability that a Markov process  starts earning rewards from state $i\in E^+$, which can either happen by $X_0=i\in E^+$ or by $X_0\in E^0$ and eventually returning to $E^+$.  
Since there in general exists the possibility of never entering $E^+$ if the process is started in $E^0$, there will
potentially be an atom (point mass) at zero of size $\pi_{d+1}=1-\vect{\pi}\vect{e}$. Hence we have proved the following:

\begin{theorem}[\cite{bladt-nielsen-2017}, p. 164] \label{TheoremNonNegRewards}
The random variable $Y$ of (\ref{eq:PH-reward}) is a mixture distribution of a point mass at 0 of size $\pi_{d+1}=1-\vect{\pi}\vect{e}$ and a phase-type distribution with
representation $\mbox{PH}_d(\vect{\pi},\mat{T}^*)$ where $\mat{T}^* = \{
t_{ij}^*:(i,j)\in E^+\}$ is given by
\begin{eqnarray*}
  t_{ij}^*= -\frac{s_{ii}}{r_i}p_{ij} \;\; \mathrm{for} \ \ i\neq j 
  \;\; \mathrm{and} \;\;
  t_{ii}^*= \frac{s_{ii}}{r_i}(1-p_{ii}).
\end{eqnarray*}
\end{theorem}
\subsection{Poisson mutations on a PH-distributed variable gives a DPH-distribution}
\label{sect:PoissonOnPH}
A result with particular relevance to coalescent theory is that a Poisson random variable with a PH-distributed rate follows a DPH-distribution. 
This result was originally stated as Theorem 3.5 in~\cite{HSB2019}. 
Here, we present a simple and alternative proof using probability generating functions, which, as we will see later, generalizes to the multivariate case.

\begin{theorem} \label{TheoremHSB35}
Consider a phase--type distributed random variable $\tau \sim {\rm PH}(\vect{\alpha},\mat{S})$. Assume $Z$ conditionally on $\tau$ is Poisson--distributed with rate $\lambda \tau$, i.e. $Z|\tau \sim {\rm Pois}(\lambda \tau)$. The resulting unconditional distribution of $Z$ is a discrete phase--type distribution 
\begin{eqnarray*}
   Z+1 \sim {\rm DPH}(\vect{\alpha},\mat{M}),
\end{eqnarray*}
where the sub--transition matrix is given by
\begin{eqnarray}
   \mat{M}=\Big(\mat{I}-\lambda^{-1} \mat{S}\Big)^{-1}. \label{eq:subtransmat}
\end{eqnarray} 
\end{theorem} 
{\bf Proof:}\\
We find
\begin{eqnarray*}
  \mathbb{E}[z^{Z+1}|\tau]
  =\sum_{s=0}^{\infty}z^{s+1}e^{-\lambda \tau}\frac{(\lambda \tau)^s}{s!}
  =ze^{-\lambda \tau}\sum_{s=0}^{\infty}\frac{(z\lambda \tau)^s}{s!}
  =ze^{-\lambda \tau} e^{\lambda \tau z}
  =ze^{\lambda \tau (z-1)},
\end{eqnarray*}
so the probability generating function is given by
\begin{eqnarray}
  \mathbb{E}[z^{Z+1}]
  =\mathbb{E}\big[ \mathbb{E}[z^{Z+1}|\tau] \big]
  =z\mathbb{E}[e^{\lambda \tau(z-1)}]
  =z\vect{\alpha}\Big(-\lambda (z-1)\mat{I}-\mat{S}\Big)^{-1}\vect{s},
  \label{PGFPoisDPH}
\end{eqnarray}
where in the second equation we used~(\ref{LaplacePH}).
So in order to show the desired result we must, according to~(\ref{DPHpgf}), show that
\begin{eqnarray}
  \Big(-\lambda(z-1)\mat{I}-\mat{S}\Big)^{-1}(-\mat{S})=
  (\mat{I}-z\mat{M})^{-1}(\mat{I}-\mat{M}).
  \label{PGFproof}
\end{eqnarray}
We have
\begin{eqnarray}
  \Big(-\lambda(z-1)\mat{I}-\mat{S}\Big)=
  \lambda\Big(\mat{I}-\lambda^{-1}\mat{S}-z\mat{I}\Big)=
  \lambda\Big( \mat{M}^{-1}-z\mat{I}\Big)=
  \lambda \mat{M}^{-1} \Big( \mat{I}-z \mat{M}\Big),
  \label{ProofEq1} 
\end{eqnarray}
so for $z=1$ this equation amounts to
\begin{eqnarray}
  -\mat{S}=
  \lambda \mat{M}^{-1} \Big( \mat{I}-\mat{M}\Big).
  \label{ProofEq2} 
\end{eqnarray}
We now obtain~(\ref{PGFproof}) by finding the inverse of~(\ref{ProofEq1}) and multiplying the result by~(\ref{ProofEq2}).
\hfill $\square$ \\
  
As a simple verification for consistency we calculate the mean number of $Z$ using the law of total expectation and Theorem~\ref{TheoremHSB35}, respectively.
The law of total expectation gives 
\begin{eqnarray}
  \Exp[Z]=\Exp\big[\Exp[Z|\tau]\big]
         =\Exp[\lambda\tau]
         =\lambda\Exp[\tau]
         =\lambda\vect{\alpha}(-\mat{S})^{-1}\vect{e}
         =\lambda \balpha \bU \be.
   \label{LawOfTotalExpectation}
\end{eqnarray}  
Using the theorem above and (\ref{DPHmean}) we get
\begin{eqnarray*}
  \Exp[Z+1] &=& \vect{\alpha} (\mat{I}-\mat{M})^{-1} \vect{e} 
           =\vect{\alpha} 
              \Big\{\mat{I}-\Big(\mat{I}-\lambda\mat{S}\Big)^{-1}\Big\}^{-1} 
              \vect{e} 
           =\vect{\alpha} 
              \Big[\mat{I}-\big\{-(\mat{I}-\lambda\mat{S})+\mat{I} \big\}^{-1} \Big]
              \vect{e} \\ 
           &=& \vect{\alpha} 
              \Big[\mat{I}-\big(\lambda\mat{S}\big)^{-1} \Big]
              \vect{e} 
           = 1+\lambda\vect{\alpha}\bU\vect{e},                   
\end{eqnarray*} 
where in the third and fourth equality we use
\begin{eqnarray}
  (\mat{A}+\mat{B})^{-1}=
  \mat{A}^{-1}-\mat{A}^{-1}(\mat{B}^{-1}+\mat{A}^{-1})^{-1}\mat{A}^{-1}.
  \label{MatrixInverse}
\end{eqnarray} 
\subsection{Multivariate phase--type (MPH) theory: Definition and Laplace transform}
In this section, we generalize the results of Section~\ref{sec:rewards} to a multivariate setting.
Let $\tau \sim \mbox{PH}_p(\vect{\alpha},\mat{S})$ and let $\{X_t\}_{t\geq 0}$ be the corresponding Markov jump process. Consider~$m$ positive reward functions 
\begin{eqnarray}
  r_j: \{ 1,\ldots,p \} \rightarrow \mathbb{R}_{+}, \;\; j=1,\ldots,m,
\end{eqnarray}
and let $\mat{R}=\{R_{ij}\}$ be the $p\times m$ matrix with entries $R_{ij}=r_j(i)$.
Hence the $j'th$ column of $\mat{R}$, $\vect{R}_{\cdot j}$, consists of $(r_j(1),\ldots,r_j(p)$.
\begin{definition}[Multivariate Phase-type (MPH) distribution] 
Let 
\begin{eqnarray}
  Y_j=\int_0^{\tau} r_j(X_t)dt=\int_0^\tau R_{X_t,j}\; dt, \;\; j=1,\ldots,m,
\end{eqnarray}
be the cumulated reward in the various states for reward function $r_j$. Then the random vector $\vect{Y}=(Y_1,\ldots,Y_m)$ is said to be multivariate phase-type distributed with parameters $\vect{\alpha}$, $\mat{S}$, and $\mat{R}$, and we write 
$\vect{Y}\sim \mbox{MPH}^{\star}(\vect{\alpha},\mat{S},\mat{R})$.
\end{definition}

The joint distribution of $\mat Y$ can be expressed in a compact form in terms of the joint Laplace transform.

\begin{theorem}[\cite{bladt-nielsen-2017} Theorem 8.1.2]
The Laplace transform for $\vect{Y}\sim \mbox{MPH}^{\star}(\vect{\alpha},\mat{S},\mat{R})$ is given by 
\begin{eqnarray}
  \mathcal{L}_{\vect{Y}}(\vect{a})=
  \mathbb{E}[e^{\vect{a}^\ast \vect{Y}}]=
  \vect{\alpha} \Big( -\mat{\Delta} (\mat{R}\vect{a})-\mat{S} \Big)^{-1} \vect{s}=
  \vect{\alpha} \Big( \mat{\Delta} (\mat{R}\vect{a})+\mat{S} \Big)^{-1} \mat{S} \vect{e},
  \label{MultLaplacePH}
\end{eqnarray} 
where $\vect{a}^\ast \vect{Y}=\sum_{j=1}^m a_j Y_j$, $\vect{s}=-\mat{S}\vect{e}$, and $\mat{\Delta}(\mat{R}\vect{a})$ is the diagonal matrix with $\mat{R}\vect{a}$ on the diagonal.
\end{theorem}

We refer to \cite{bladt-nielsen-2017} for a proof of the Theorem.

Of special interest are means, variances and covariances between elements of $\vect{Y}$. Let $\vect{R}_{\cdot i}$ denote the $i$th column of $\mat{R}$ and recall $\mat{U}=(-\mat{S})^{-1}$ is the Green matrix. Then we have  
\begin{eqnarray}
  \Exp [Y_i] &=& \vect{\alpha} \mat{U}\vect{R}_{\cdot i},
   \label{eq:gen-mean}
\end{eqnarray}
and for $i,j = 1,\ldots,m$ we have
\begin{eqnarray}
  \Exp[Y_i Y_j]= \vect{\alpha} 
    \mat{U}\mat{\Delta}(\mat{R}_{\cdot i})\mat{U}\mat{R}_{\cdot j}  + 
    \vect{\alpha} \mat{U}\mat{\Delta}(\mat{R}_{\cdot j})\mat{U}\mat{R}_{\cdot i}.
 \label{eq:gen-cross-moment}
\end{eqnarray}
\subsection{Poisson mutations on a MPH*-distributed random variable}
\label{sect:PoissonOnMPH}
In Theorem~\ref{TheoremHSB35} we saw that adding Poisson mutations on a PH-distributed random variable (which may arise via reward transformation) resulted in a DPH-distributed random variable whose sub-transition matrix is given in terms of the underlying sub-intensity matrix (recall equation~\eqref{eq:subtransmat}).
In this section we give the corresponding multivariate result of Poisson mutations on a
MPH*-distributed random variable. We calculate the probability generating function (PGF) of the resulting distribution, which, unlike in the univariate case, does not belong to a class of distributions considered so far. 

\begin{theorem} \label{multPGFtheorem}
Assume $\vect{Y}=(Y_1,\ldots,Y_m)\sim \mbox{MPH}^{*}(\vect{\alpha},\mat{S},\mat{R})$ and assume the entries in $\vect{Z}=(Z_1,\ldots,Z_m)$ conditional on $\vect{Y}$ are independent Poisson distributed with rates $\lambda \vect{Y}$, i.e. 
$Z_j|Y_j \sim {\rm Pois}(\lambda Y_j)$.
The PGF for~$\vect{Z}$ is given by
\begin{eqnarray}
   \varphi(\bz) \eqdef \mathbb{E}[z_1^{Z_1}\cdots z_m^{Z_m}] =
  \vect{\alpha}
    \big( \mat{\Delta}(\mat{R}\lambda(\vect{z}-\vect{e}))+\mat{S} \big)^{-1}\mat{S}
  \vect{e},
  \label{multPGF}
\end{eqnarray}
\end{theorem}
{\bf Proof:} A direct calculation gives the result
\begin{eqnarray}
   \varphi(\bz) & \eqdef &  
  \mathbb{E}[z_1^{Z_1}\cdots z_m^{Z_m}] \\
  &=&  
  \mathbb{E}\Big[\mathbb{E}[z_1^{Z_1}\cdots z_m^{Z_m}|\vect{Y}]\Big] 
  \nonumber \\
  &=&
  \mathbb{E} 
  \Big[
  \mathbb{E}[z_1^{Z_1}|Y_1] \cdots \mathbb{E}[z_m^{Z_m}|Y_m]
  \Big] 
  \nonumber \\
  &=&
  \mathbb{E} 
  \Big[
  e^{\lambda Y_1 (z_1-1)} \cdots e^{\lambda Y_m (z_m-1)}
  \Big] 
  \nonumber \\
  &=&
  \mathbb{E} 
  \Big[
   e^{\lambda Y_1 (z_1-1)+ \cdots + \lambda Y_m (z_m-1)}
  \Big] 
  \nonumber \\
  &=&
  \mathbb{E} 
  \Big[
   e^{\lambda(\vect{z}-\vect{e})^{\ast} \vect{Y}}
  \Big] 
  \nonumber \\
  &=&
  \vect{\alpha}
  (\mat{\Delta}(\mat{R}\lambda(\vect{z}-\vect{e}))+\mat{S})^{-1}\mat{S}\vect{e},
\end{eqnarray}
where in the last equation we used (\ref{MultLaplacePH}).
\hfill $\square$ \\

We get the PGF for $Z_i$ by setting all entries of $\vect{z}$ to one except for entry $i$ and obtain 
\begin{eqnarray*}
  \varphi_i(z_i)&=&
  \varphi\Big( (z_i-1)\be_i+\be \Big)=
  \vect{\alpha}
  (-\lambda(z_i-1) \mat{\Delta}(\mat{R}_{\cdot i})-\mat{S})^{-1}
  \vect{s}, 
\end{eqnarray*} 
where we assume all entries in $\mat{R}_{\cdot i}$ are positive.
From (\ref{PGFPoisDPH}) we observe that $Z_i+1$ is 
DPH-distributed with initial probability vector $\vect{\alpha}$ and sub-transition matrix $(\mat{I}-\frac{1}{\lambda} \mat{\Delta}(\mat{R}_{\cdot i})^{-1} \mat{S})^{-1}$.
This result is of course also an immediate consequence of $Y_i$ following a ${\rm PH}(\balpha,\bS,\bR_{i.})$ distribution, and $Z_i|Y_i$ following a Poisson distribution with rate $\lambda Y_i$.

We now consider the joint distribution of $Z_i$ and $Z_j$. We get the PGF for $(Z_i,Z_j)$ by setting all entries of $\bZ$ to one except for entry~$i$ and~$j$ and obtain
\begin{align}
  \varphi_{ij}(z_i,z_j)=
  \varphi\Big( (z_i-1)\be_i+(z_j-1)\be_j+\be \Big)=
  \vect{\alpha}
  \Big( -\lambda(z_i-1) \mat{\Delta}(\mat{R}_{\cdot i})-
  \lambda(z_j-1) \mat{\Delta}(\mat{R}_{\cdot j})-
  \mat{S} \Big)^{-1}
  \vect{s}.
\end{align} 
We now find the mean of $Z_iZ_j$ by differentiating with respect to $z_i$ and $z_j$ and evaluating in (1,1). From (\ref{DerivSumInverse}) we get
\begin{eqnarray*}
\Exp[Z_iZ_j] &=&
\vect{\alpha} \Big(
  \lambda \bS^{-1} 
  \mat{\Delta}(\mat{R}_{\cdot i}) \bS^{-1} \mat{\Delta}(\mat{R}_{\cdot j}) 
  \bS^{-1}+
  \lambda \bS^{-1} 
  \mat{\Delta}(\mat{R}_{\cdot j}) \bS^{-1} \mat{\Delta}(\mat{R}_{\cdot i}) 
  \bS^{-1}
\Big) \mat{S} \vect{e} \nonumber \\
&=&
 \vect{\alpha} 
 \lambda
  \bS^{-1} 
  \mat{\Delta}(\mat{R}_{\cdot i}) \bS^{-1} \mat{\Delta}(\mat{R}_{\cdot j}) 
  \vect{e}+
  \vect{\alpha} 
  \lambda \bS^{-1} 
  \mat{\Delta}(\mat{R}_{\cdot j}) \bS^{-1} \mat{\Delta}(\mat{R}_{\cdot i}) 
  \vect{e} \\
  &=& 
  \lambda \big( \vect{\alpha} 
  \bU 
  \mat{\Delta}(\mat{R}_{\cdot i}) \bU \mat{R}_{\cdot j} 
  +
  \vect{\alpha} 
  \bU 
  \mat{\Delta}(\mat{R}_{\cdot j}) \bS^{-1} \mat{R}_{\cdot i} \big), 
\end{eqnarray*}  
and we note that this expression can also be obtained using a conditional argument and equation~(\ref{eq:gen-cross-moment}).

We finally note, for future reference, that the law of total variance gives 
\begin{eqnarray}
  \Var(\bZ)=
  \Exp[ \Cov(\bZ \mid \bY) ]+ \Cov( \Exp[\bZ \mid \bY] ) 
  = \lambda \mat{\Delta}(\bmu) + \lambda^2 \bSigma, 
  \label{eq:covmatdecomp}
\end{eqnarray}
where $\mat{\mu}$ and $\mat{\Sigma}$ denotes the mean and covariance matrix of $\bY$ given by~(\ref{eq:gen-mean}) and (\ref{eq:gen-cross-moment}).
\section{Phase--type distributions in coalescent theory: \\
The block-counting process and zero--one coefficients of the SFS} \label{01coefficients}
\begin{figure}[h]
\centering
\includegraphics[scale=1.2]{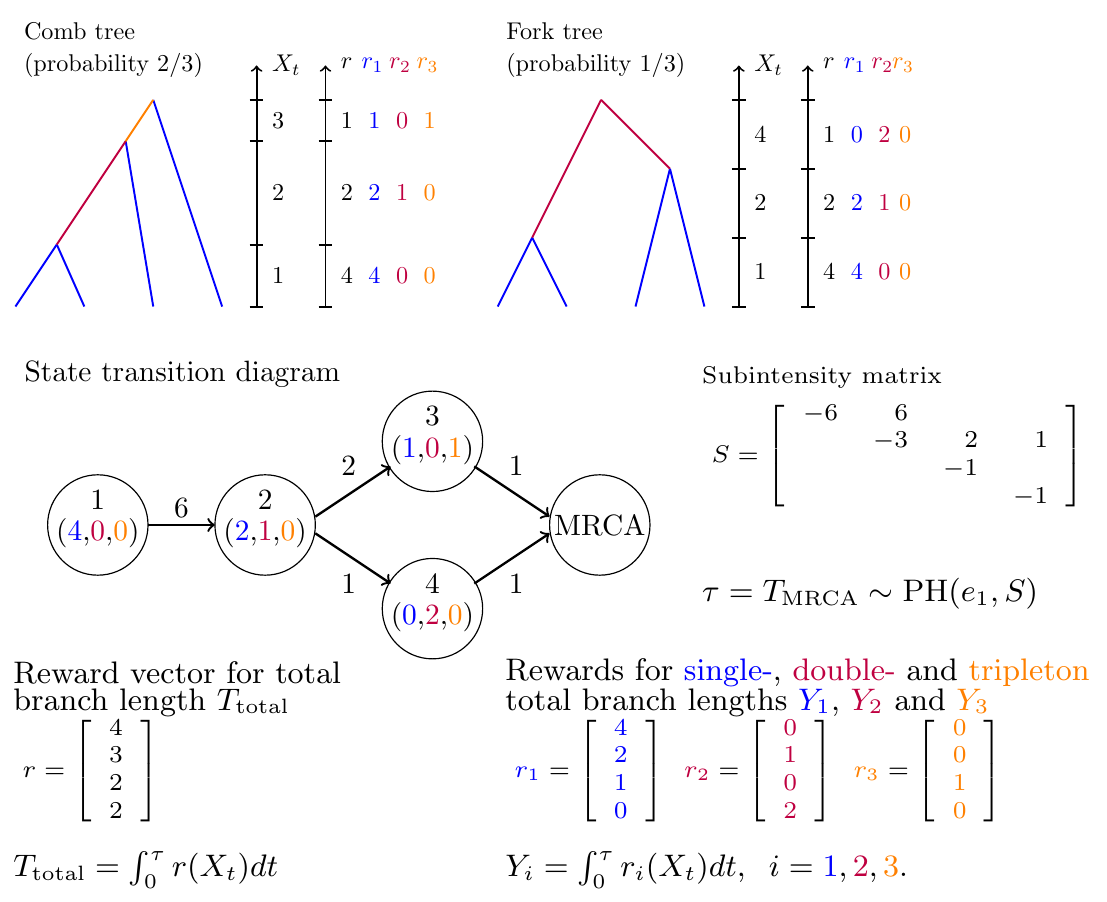}
\caption{Illustration of the block-counting process and construction of the phase--type distribution for total branch lengths with one, two or three descendants in a sample of size $n=4$.}
\label{MPHfigure}
\end{figure}

The central element, which will allows us to use the phase--type theory 
introduced in the previous section to model the site frequency spectrum, is the so-called \textit{block-counting process}.
The block-counting process was introduced in \cite{HSB2019} and tracks the number of branches in a coalescent tree, which has 
$i$ descendants in the sample, $i=1,\dots,n-1$, and is illustrated in Figure~\ref{MPHfigure} for the case $n=4$. 
We use $\{X_t\}_{t\geq 0}$ to denote the state of the process. The starting state (state~1) is indexed by $(4,0,0)$, which corresponds to four \textit{'singleton branches'}, zero \textit{'doubleton branches'} and zero \textit{'tripleton branches'}. Here, a \textit{'$i$-ton branch'} is a branch with $i$ present-day descendants. The coalescent rate is ${4 \choose 2}=6$ before coalescent, and the next state (state~2) is indexed by $(2,1,0)$ and consists of two singletons, one doubleton and zero tripleton branches. Two types of coalescent events are possible from state 2: A coalescent of one of the two singleton branches and the doubleton branch or a coalescent of the two singleton branches. The first event happens with rate~$2 \cdot 1=2$ and results in one singleton, zero doubleton and one tripleton branch (state~3; indexed by $(1,0,1)$). The second event happens with rate~${2 \choose 2}=1$ and results in state~4 (indexed by $(0,2,0)$). The first event results in a comb tree, and the second event in a fork tree. From state~3 or state~4 the two branches coalesce with a rate of~1, and after the event we are in the absorbing state, which corresponds to the most recent common ancestor (MRCA). 

Now, we observe that the time to the most recent common ancestor $\tau \eqdef T_{\rm MRCA}$ has a phase--type distribution $\mbox{PH}_{4}(\vect{e}_1,\mat{S})$. Furthermore, by considering the number of branches in each state, we see that the total branch length can be expressed as  
\begin{equation}
Y = \int_0^\tau r_i \indi{X_t = i } \dd t \, , \label{eq:rewardproc2}
\end{equation}
where $\br = (r_1,r_2,r_3,r_4) = \left(4,3,2,2\right)$ and from (\ref{PosRewards}) we get that the total branch length $T_{\rm total}$ has a phase--type distribution $\mbox{PH}_{4}\big(\vect{e}_1,\mat{\Delta}^{-1}(\br)\mat{S}\big)$. 
Finally, it follows from Theorem \ref{TheoremHSB35}, that the total number of segregating sites plus one, $\xi_{\rm total}=\xi_1+\ldots+\xi_{n-1} + 1$, is discretely phase--type distributed with representation 
${\rm DPH}(\be_1,\bM)$ where
$$
\bM = \left( \bI - \frac{2}{\theta} \bDelta^{-1}(\br)\mat{S} \right)^{-1}.
$$
An analytical expression for the distribution of the number of segregating sites can also be found in \cite[\S 4.1.1]{wakeley2008coalescent}.

Next, suppose we wish to determine the distribution of the number of singletons $\xi_1$. A mutation is a singleton in our sample when it occurs on a singleton-branch. The number of such branches is recorded by the first entry in the block counting process, and we can find the total branch length of all singleton-branches using~(\ref{eq:rewardproc2}) with reward vector $\br = \left(4,2,1,0\right)$ (see Figure~ \ref{MPHfigure}), so that by~Theorem~\ref{TheoremNonNegRewards} this random variable has a PH-representation, and we can again use Theorem~\ref{TheoremHSB35} to conclude that $\xi_1+1$ is DPH-distributed. 
Note that $\br$ is the first entry in the vector representation of the state-space in the state transition diagram in Figure~\ref{MPHfigure}.
Actually, a consequence of Theorem~\ref{TheoremHSB35} is that the singleton, doubleton etc. branch lengths are a mixture of a point mass at 0 and a phase--type distribution. 
The argument above extends to any 0--1--weighted version of the site frequency spectrum, which is stated as Theorem \ref{01Theorem} below.

\begin{figure}[h]
\centering
\includegraphics[scale=1]{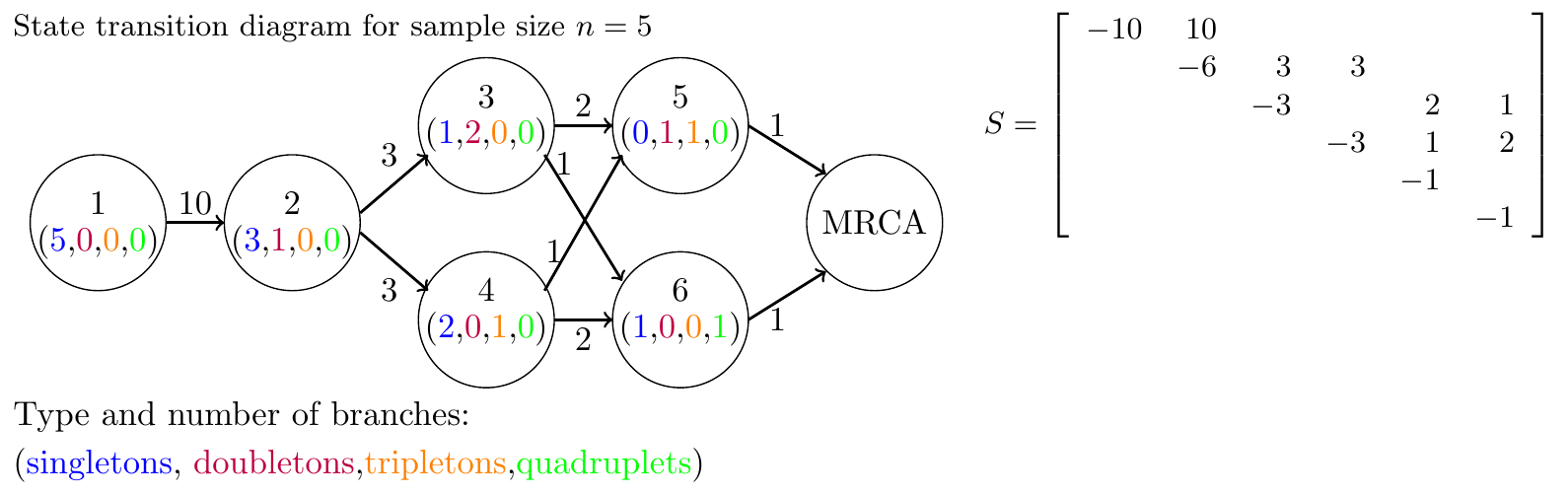}
\caption{State transition diagram, sub--intensity matrix and type and number of branches for each state for the construction of the reward vector for the total branch lengths with one, two three or four descendants in a sample of size $n=5$.}
\label{Kingman5Figure}
\end{figure}

In Figure~\ref{Kingman5Figure} we show the state transition diagram and subintensity matrix for the block counting process for $n=5$, and we
now turn to the general formulation of the block-counting process. For a general sample size~$n$ the states are represented by the vector $\vect{a}=(a_1,a_2,...,a_n)$ where $a_i$ denotes the number of branches with $i$ descendants. The state-space is thus given by 
\begin{eqnarray*}
  \Big\{\vect a=(a_1,\dots,a_{n-1}) \in\mathbb Z_+^{n-1} \; :
         \; \sum_{i=1}^{n-1} ia_i=n\Big\}.
\end{eqnarray*}
For the standard coalescent the possible transitions are 
\[ (a_1,\dots,a_{n-1}) \rightarrow (a_1,\dots,a_i-1,\dots, a_j-1\dots, a_{i+j}+1,\dots,a_{n-1}) \]
with rate $a_ia_j$ for $a_i,a_j\geq 1$, and
 \[ (a_1,\dots,a_{n-1}) \rightarrow (a_1,\dots,a_i-2,\dots, a_{2i}+1,\dots,a_{n-1})  \]
with rate ${a_i \choose 2}$ for $a_i\geq 2$. For example, using the enumeration of the states given in Figure~\ref{MPHfigure} for~$n=4$ and Figure~\ref{Kingman5Figure} for~$n=5$ the state-space for these two cases are given by the rows in the matrices
\begin{eqnarray}
\bA_4=
\begin{pNiceMatrix}[first-col]
1 & 4 & 0 & 0 \\
2 & 2 & 1 & 0 \\
3 & 1 & 0 & 1 \\
4 & 0 & 2 & 0 
\end{pNiceMatrix}
\;\;\; {\rm and} \;\;\;
\bA_5=
\begin{pNiceMatrix}[first-col]
1 & 5 & 0 & 0 & 0 \\
2 & 3 & 1 & 0 & 0 \\
3 & 1 & 2 & 0 & 0 \\
4 & 2 & 0 & 1 & 0 \\
5 & 0 & 1 & 1 & 0 \\
6 & 1 & 0 & 0 & 1 
\end{pNiceMatrix}
\label{eq:statespaces}
\end{eqnarray}
An algorithm for generating the state-space and the corresponding rate matrix is given as Algorithm 4.2 in \cite{HSB2019}, and the algorithm is implemented in the \texttt{phasty} package. In the following we suppress the dependence on $n$ in $\mat{A}_n$ and just write $\mat{A}$.

We remark that the size of the state space for the block counting process is given by the so-called partition function $p(n)$ minus one. For example $p(4)=5$ because~4 can be partitioned as $1+1+1+1$, $1+1+2$, $1+3$, $2+2$ and $4$.
A few selected values of the partition function are $p(5)=7$, $p(10)=42$, $p(15)=176$, $p(20)=627$, $p(25)=1958$ and $p(30)=5604$. We observe that the partition function grows very fast in $n$. Actually, \cite{HardyRamanujan1918} showed that the partition function can be approximated by 
\begin{eqnarray*}
  p(n) \approx \frac{1}{4n\sqrt{3}} e^{\pi \sqrt{2n/3} }. 
\end{eqnarray*} 
The fast increase in the size of the state space means that we are limited to sample sizes smaller than~$n=30$.  

The reasoning above concerning the distribution of singletons extends to the multivariate case. For example, if we wish to consider the simultaneous
distribution of $(\xi_1,\xi_2)$, the number of singletons \textit{and} the number of doubletons, we first need to consider the simultaneous
distribution of the  total branch length of all singleton-branches \textit{and} the  total branch length of all doubleton-branches, and this can be
obtained via a reward transformation of the block-counting process with \emph{two} reward functions, namely $\br_1 = (4,2,1,0)^\ast$ and $\br_2 = (0,1,0,2)^\ast$, which correspond to columns $1$ and $2$ of the state-space matrix from \eqref{eq:statespaces}. The distribution of $(\xi_1,\xi_2)$
can be found using the results from Section~\ref{sect:PoissonOnMPH}. We state these observations as a theorem for the entire site frequency spectrum~$\bxi$.

\begin{theorem} \label{thm:sfs}
Let $\bA$ be the state-space and $\bT$ the corresponding sub-transition matrix of the block-counting process. The singleton, doubleton etc. branch length vector $\bY=(Y_1,\ldots,Y_{n-1})$ is multivariate phase--type distributed
\begin{eqnarray*}
  \bY \sim {\rm MPH}^{\star}(\be_1,\bT,\bA).
\end{eqnarray*} 
Furthermore, conditionally on the the branch length vector, the entries $\xi_i$, $i=1,\ldots,n-1$, in the site frequency spectrum are mutually independent and Poisson distributed   
\begin{align}
  \xi_i \mid \bY \sim \Po\Big(Y_i \frac{\theta}{2}\Big),   
\label{eq:sfsdist2}
\end{align}
where $\theta$ is the mutation rate.
The unconditional distribution of the site frequency spectrum has PGF
\begin{eqnarray}
   \varphi_\bxi(\bz)=  
   \vect{e}_1  
   (-\triangle(\mat{A}\lambda(\vect{z}-\vect{e}))-\mat{T})^{-1}
   \vect{t},
  \label{eq:pgfSFS}
\end{eqnarray}
where $\vect{t}=-\mat{T}\vect{e}$.
\end{theorem}
Here, \eqref{eq:pgfSFS} follows immediately from \eqref{eq:sfsdist2} using 
Theorem~\ref{multPGFtheorem} in Section~\ref{sect:PoissonOnMPH}. This is, to our knowledge, the first analytic description of the simultaneous distribution of the site frequency spectrum. 

As described in the introduction, weighted versions of the SFS are of considerable interest, and while the PGFs of the corresponding 
distributions are readily available from Theorem~\ref{thm:sfs}, the resulting expressions in the case of general coefficients are difficult to work with
from a numerical point of view. We return to this point
at the end of Section \ref{integerCoefficients}, but will now continue by restricting
the kinds of weights we consider, which leads to 
the tractable class of DPH-distributions Our first result in this direction is the following:

\begin{theorem} \label{01Theorem}
  Any 0--1--weighted version of the SFS is DPH--distributed.
\end{theorem}
{\bf Proof:}\\
The result is an immediate consequence of Theorem~\ref{TheoremNonNegRewards} and Theorem~\ref{TheoremHSB35} when applied to the block-counting process. In particular, in Theorem~\ref{TheoremNonNegRewards}, we showed that a non--negative weighted version of the states are PH--distributed (possible defective). By choosing the weights appropriately we obtain the total branch of all $i$-ton-branches desired (note that e.g. including both singletons and doubletons is simply a question of adding the respective reward vectors). In Theorem~\ref{TheoremHSB35} we showed that Poisson mutations on a PH--distributed random variable is DPH--distributed.
\hfill $\square$ \\

Examples include the number of singletons $\xi_1$, doubletons $\xi_2$, and in general the number of $i$-tons $\xi_i$. Furthermore, the total number of segregating sites $\xi_{\rm total}=\xi_1+\ldots+\xi_{n-1}$, the tail statistics $\xi_{i+}=\xi_i+\ldots+\xi_{n-1}$ for $i=1,\ldots,n-1,$ (\cite{Koskela2018}) and the entries in the folded SFS e.g. $\eta_1=\xi_1+\xi_{n-1}$ also belong to this class. 

\begin{figure}[h]
\centering
\includegraphics[scale=0.33]{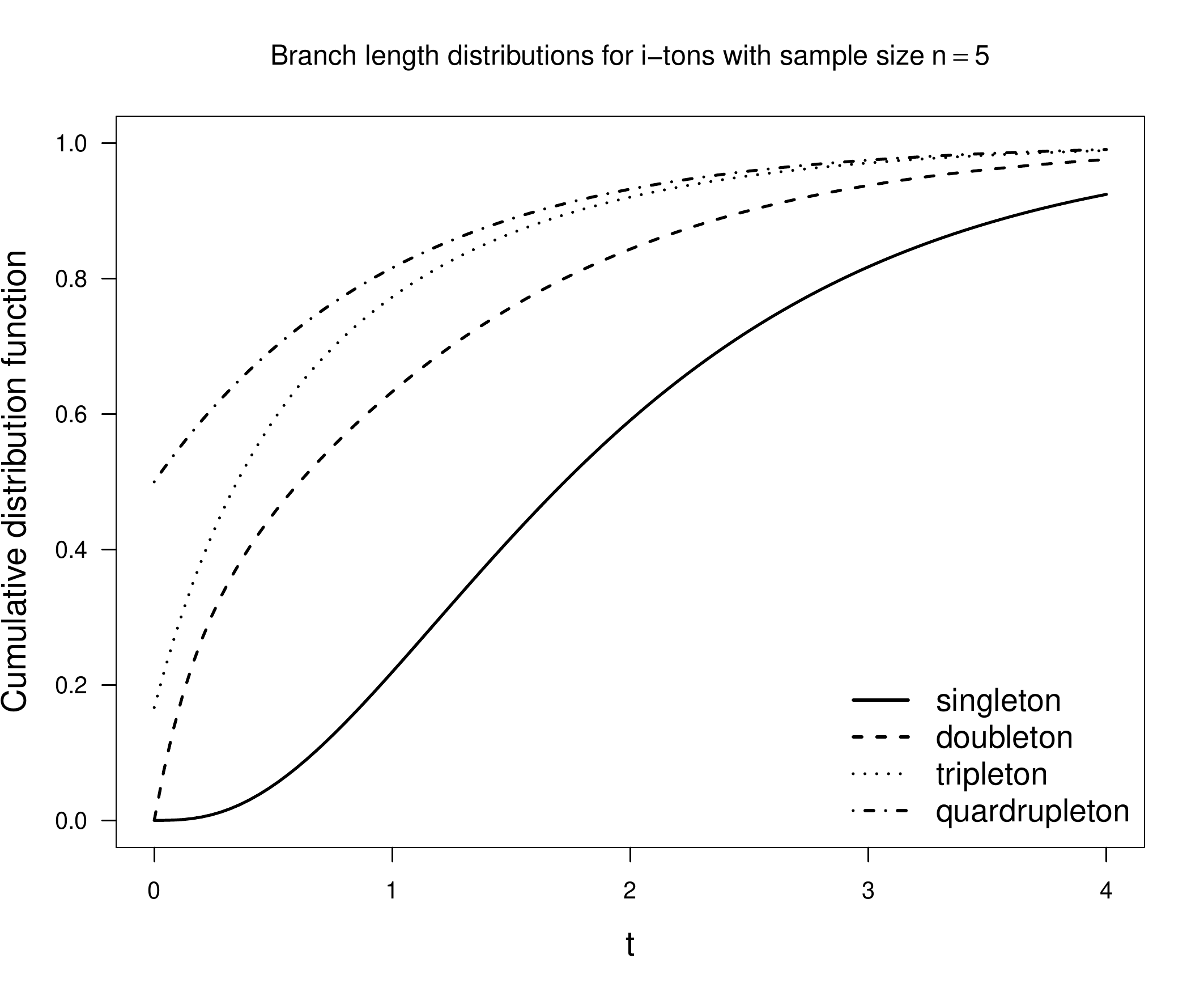}
\includegraphics[scale=0.33]{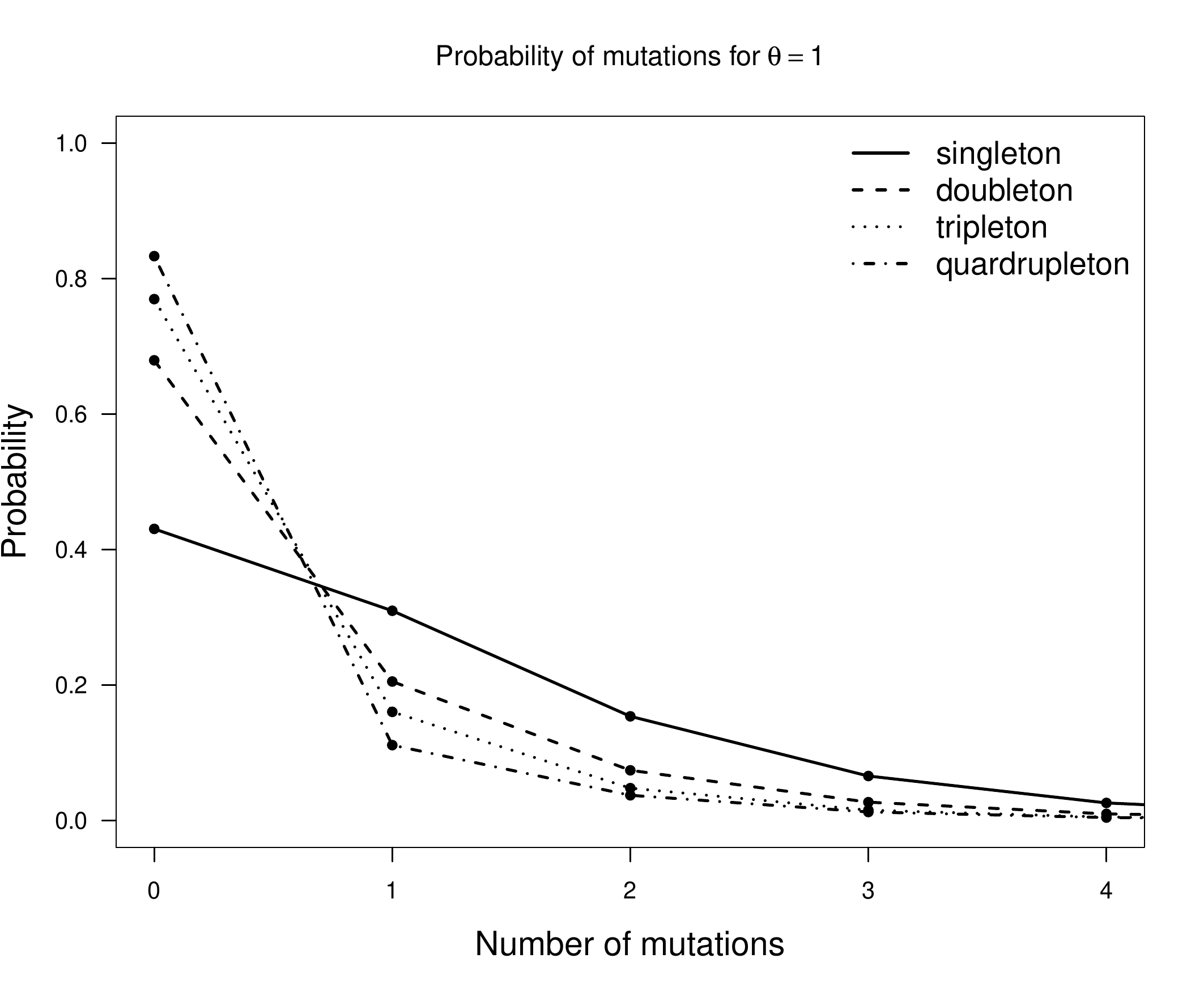}
\caption{Left: Branch length distributions for $i$-tons, $i=1,2,3,4$, when $n=5$. 
Right: Corresponding probabilities of number of mutations in the site frequency spectrum for $\theta=1$. The mean of the distributions are given by the classical result $\theta/i$ (e.g. \cite{wakeley2008coalescent}~page~102-103); also recall equation~(\ref{meanSFS})).}
\label{Kingman5Figurev2}
\end{figure}

The utility of Theorem~\ref{01Theorem} is apparent from Figure~\ref{Kingman5Figurev2}, where we have used the \texttt{phasty} package to compute the marginal distribution of singleton, doubleton, tripleton and quadrupleton branch lengths with $n=5$. The tripleton and quadrupleton branch lengths are defective phase--type; the singleton and doubleton branch lengths do not have a defect. In the right plot in Figure~\ref{Kingman5Figure} we show the marginal distribution for the number of mutations. For all four types of mutations, the distribution is discrete phase--type. 

\section{Non--negative integer--valued coefficients} \label{integerCoefficients}
Recall the estimators of the mutation rate $\theta$ from 
eqn. (\ref{PairwiseDifferenceEstimator}) and eqn. (\ref{HLestimators}) 
\begin{eqnarray*}
  \frac{n(n-1)}{2} \hat{\theta}_{\pi}=\sum_{i=1}^{n-1} i(n-i)\xi_i, \;\;\;\; 
  \frac{n(n-1)}{2} \hat{\theta}_{\rm H}=\sum_{i=1}^{n-1} i^2 \xi_i, \;\;\;\;
  {\rm and} \;\;\;\;
  (n-1) \hat{\theta}_{\rm L}=\sum_{i=1}^{n-1} i \xi_i,
\end{eqnarray*}
and note that $\hat{\theta}_{\pi}, \hat{\theta}_{\rm H}$ and $\hat{\theta}_{\rm L}$ are (up to the scaling constants $n(n-1)/2$ or $(n-1)$) examples of non--negative and integer--valued weighted functions of the SFS. In this section we demonstrate the following result:
\begin{theorem}
  Any non--negative integer--weighted version of the SFS is DPH--distributed.
  \label{NonNegIntegerSFSTheorem}
\end{theorem} 

\begin{figure}[h]
\centering
\includegraphics[scale=1.2]{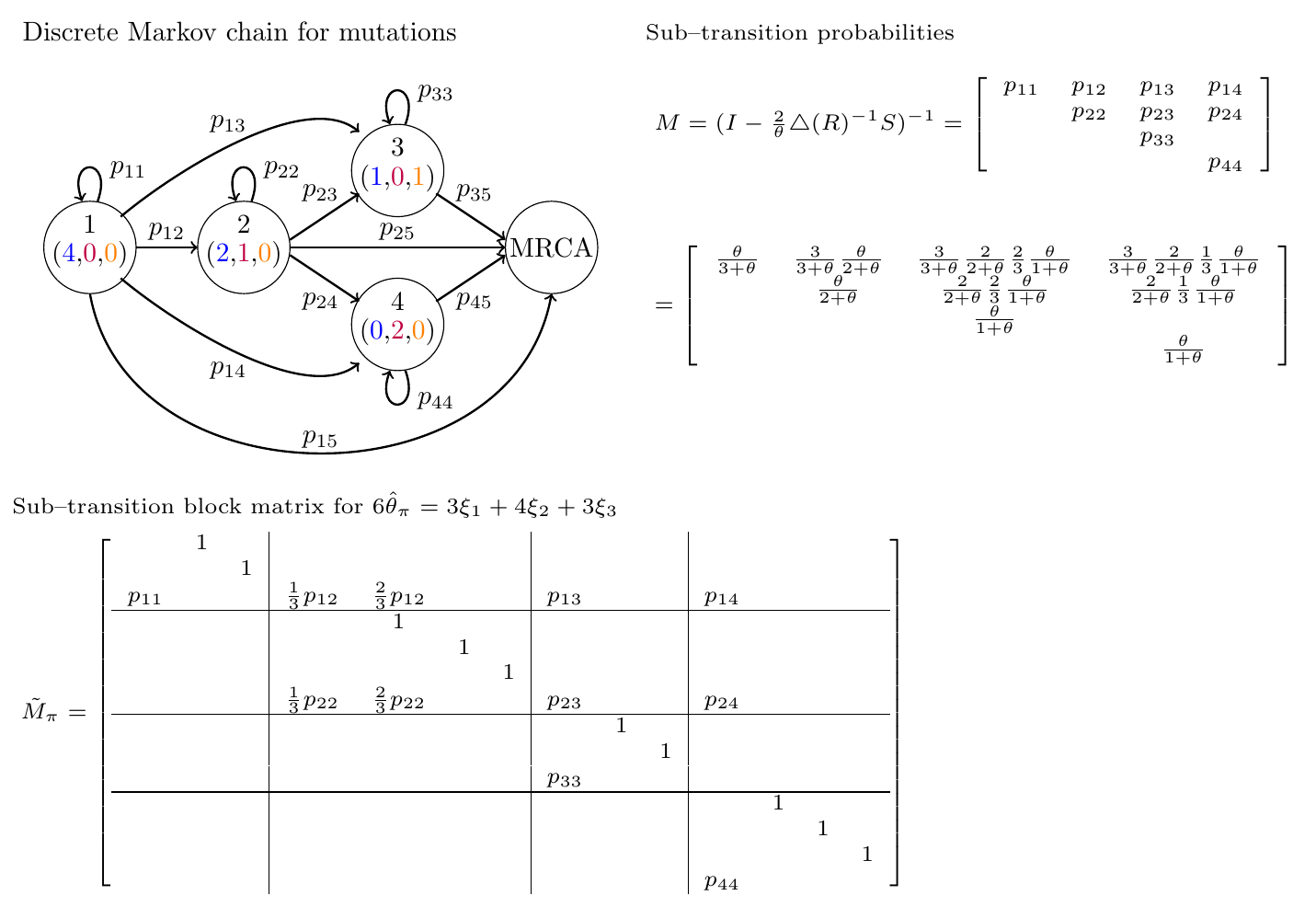}
\caption{Construction of the discrete phase--type distribution for calculating the distribution of $3\xi_1+4\xi_2+3\xi_3$ in a sample of size $n=4$. Top left: Each transition corresponds to a mutation in a state except when the transition is to the MRCA. Top right: The transition probabilities are determined by the rate matrix $S$ from Figure~\ref{MPHfigure} and reward matrix $\bA_4$ from eqn.~(\ref{eq:statespaces}). Bottom: The distribution of $1+3\xi_1+4\xi_2+3\xi_3$ is discrete phase--type distributed with initial distribution $\vect{e}_1$ and sub--transition matrix $\tilde{\mat{M}}_{\pi}$.}
\label{BlockDPHpairwiseFigure}
\end{figure}

Perhaps the easiest procedure for showing Theorem~\ref{NonNegIntegerSFSTheorem} is by means of construction. In Figure~\ref{BlockDPHpairwiseFigure} we illustrate how the distribution of the pairwise estimator can be calculated for sample size $n=4$. The four possible states are given in the top left corner of the figure, and the corresponding rates between the states are given in Figure~\ref{MPHfigure}. The number of mutations in each state is, according to  Theorem~\ref{TheoremHSB35}, given by the upper right corner with rate matrix $\bS$ given in Figure~\ref{MPHfigure} and reward matrix $\bA_4$ from eqn.~(\ref{eq:statespaces}). In the variable $3\xi_1+4\xi_2+3\xi_3$ a singleton or tripleton mutation contributes by three. Therefore a mutation in state~1 (only singleton mutations possible) or state~3 (only singleton or tripleton mutations can occur) always result in three transitions in the discrete Markov chain. This property is achieved by the $3\times 3$ block constructions for these two states in the sub--transition matrix $\tilde{\mat{M}}_{\pi}$ in the bottom of Figure~\ref{BlockDPHpairwiseFigure}. Doubleton mutations contribute by four in the variable, and all mutations in state~4 are doubletons. Four transitions in state four is achieved by the $4\times 4$ block for this state. Finally, in state~2 we have 2 singleton branches and 1 doubleton branch. Therefore a mutation in this state contributes three transitions with probability $2/3$ and four transitions with probability $1/3$. These probabilities and number of transitions are obtained by the block construction for state~2. In conclusion we have that $1+3\xi_1+4\xi_2+3\xi_3$ is discrete phase--type distributed with initial distribution $\vect{e}_1$ and sub--transition matrix $\tilde{\mat{M}}$.    

For $n=4$ we get $3\hat{\theta}_L=\xi_1+2\xi_2+3\xi_3$. In this case $1+\xi_1+2\xi_2+3\xi_3$ is phase--type distributed with initial distribution $\vect{e}_1$ and sub--transition $8\times 8$ block matrix
\begin{eqnarray}
\tilde{\mat{M}}_{\rm L}=
  \left[
  \begin{array}{c|cc|ccc|cc}
     p_{11} & \frac{1}{3} p_{12} & \frac{2}{3}p_{12} 
            & \frac{1}{2} p_{13} & & \frac{1}{2} p_{13} & p_{14} & \\ \hline
            &        & 1 & & &  \\
            & \frac{1}{3} p_{22} & \frac{2}{3}p_{22} 
            & \frac{1}{2}p_{23} & & \frac{1}{2}p_{23} & p_{24} & \\ \hline
            & & & & 1 & \\
            & & & & & 1 & \\
     & & & \frac{1}{2}p_{33} & & \frac{1}{2}p_{33} & \\ \hline
     & & & & & & & 1 \\
     & & & & &  & p_{44}  \\
  \end{array}
  \right].
  \label{ML}
\end{eqnarray}
For $n=4$ we have $6\hat{\theta}_H=\xi_1+4\xi_2+9\xi_3$, and we observe that $1+\xi_1+4\xi_2+9\xi_3$ is phase--type distributed with initial distribution $\vect{e}_1$ and sub--transition $18\times 18$ block matrix
\begin{eqnarray}
\tilde{\mat{M}}_{\rm H}=
  \left[
  \begin{array}{c|cccc|ccccccccc|cccc}
   p_{11} & \frac{1}{3} p_{12} &&& \frac{2}{3} p_{12} & 
   \frac{1}{2} p_{13} &&&&&&&& \frac{1}{2} p_{13} & p_{14} &&& \\ \hline
   && 1 &&&&&&&&&&&&&&& \\ 
   &&& 1 &&&&&&&&&&&&&& \\  
   &&&& 1 &&&&&&&&&&&&& \\
   & \frac{1}{3} p_{22} &&& \frac{2}{3}p_{22} &
   \frac{1}{2}p_{23} &&&&&&&& \frac{1}{2}p_{23} & p_{24} &&& \\ \hline
   &&&&&& 1 &&&&&&&&&&& \\
   &&&&&&& 1 &&&&&&&&&& \\
   &&&&&&&& 1 &&&&&&&&& \\
   &&&&&&&&& 1 &&&&&&&& \\
   &&&&&&&&&& 1 &&&&&&& \\
   &&&&&&&&&&& 1 &&&&&& \\
   &&&&&&&&&&&& 1 &&&&& \\
   &&&&&&&&&&&&& 1 &&&& \\
   &&&&& \frac{1}{2}p_{33} &&&&&&&& \frac{1}{2}p_{33} \\ \hline
   &&&&&&&&&&&&&&& 1 && \\
   &&&&&&&&&&&&&&&& 1 & \\
   &&&&&&&&&&&&&&&&& 1  \\
   &&&&&&&&&&&&&& p_{44}  \\
  \end{array}
  \right].
  \label{MH}
\end{eqnarray}

\begin{figure}[h]
  \centering
  \includegraphics[scale=0.33]{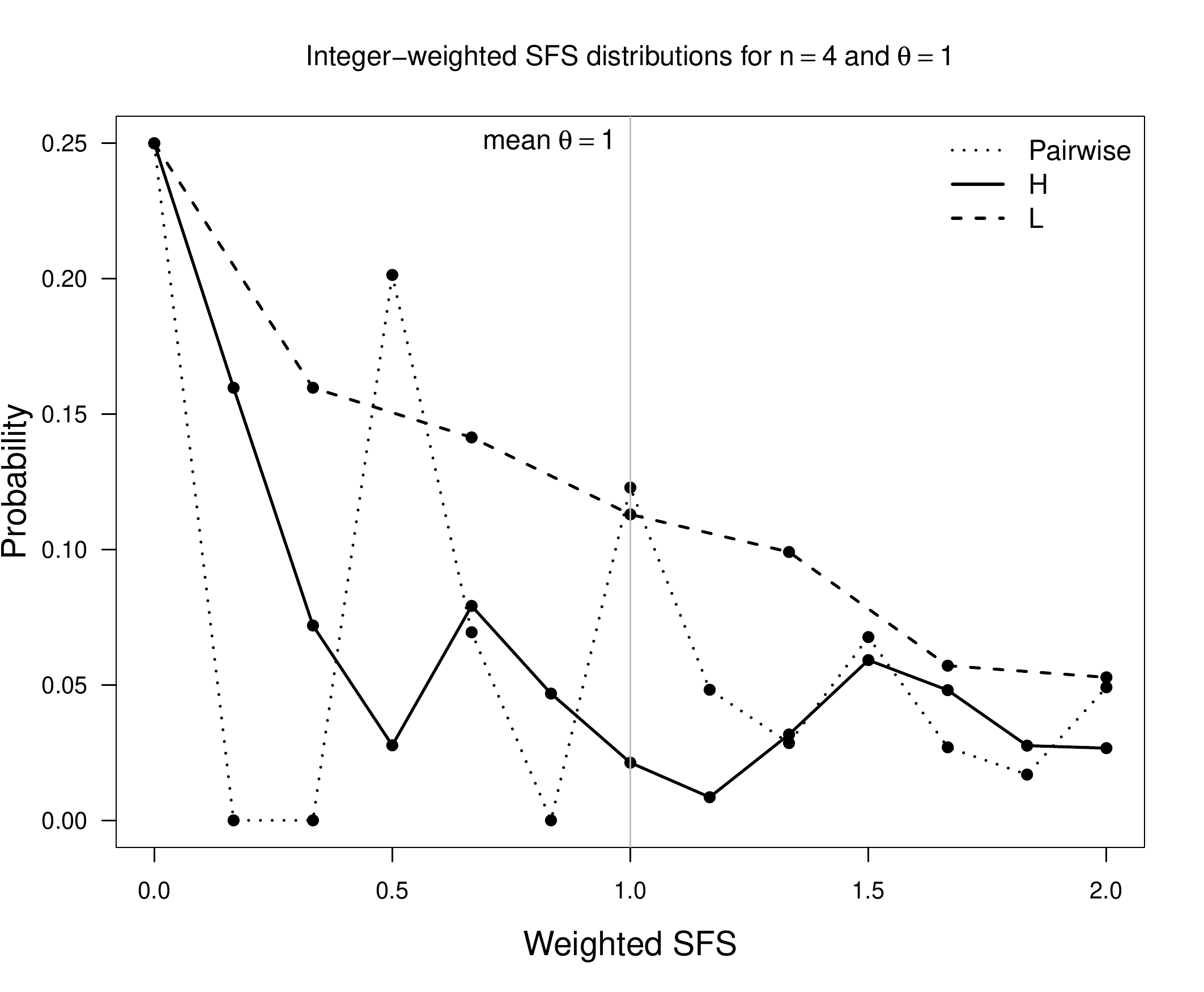}
  \includegraphics[scale=0.33]{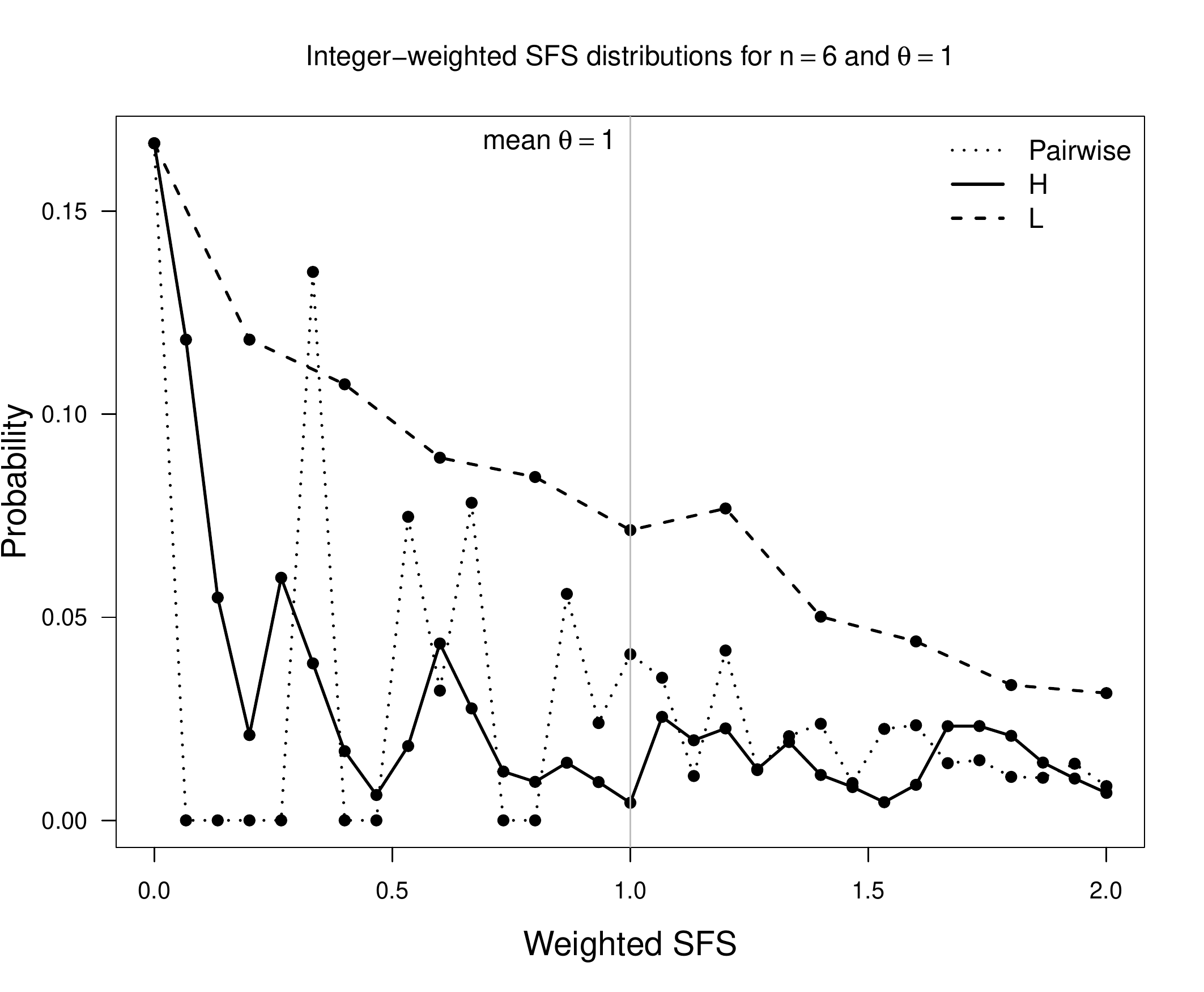}
  \caption{Left: The distribution of three integer--weighted estimators for $\theta$ ($\hat{\theta}_{\pi}, \hat{\theta}_H$ and $\hat{\theta}_L$) for $n=4$. Right: The same three distributions for $n=6$.}
  \label{IntegerWeightedFigure}
\end{figure}

In Figure~\ref{IntegerWeightedFigure} we show the distributions of $\hat{\theta}_{\pi}$, $\hat{\theta}_{\rm L}$ and $\hat{\theta}_{\rm H}$. In both plots we have $\theta=1$. In the left plot we have $n=4$, and in the right plot we have $n=6$. We see that the support for $\hat{\theta}_{\pi}$ is rather curios: for $n=4$ the support is $j/6$ for $j$ a non--negative integer but with $j\notin\{1,2,5\}$. For $n=6$ we have 
$15\hat{\theta}_{\pi}=5\xi_1+8\xi_2+9\xi_3+8\xi_4+5\xi_5$, and the support is therefore $j/15$ with $j$ a non--negative integer but with $j\notin\{1,2,3,4,6,7,11,12\}$.

The general situation for positive integer--valued weights is formulated in  Appendix~\ref{GenSectDPH}. Basically we provide a mathematical description of the construction from the three examples above. 
We also refer to the implementation in the accompanying R package.
Finally we remark that if one or more states have a zero reward, then these states can be handled using a similar construction as in (\ref{ZeroConstruction}).  
\section{General coefficients and inversion of the characteristic function} \label{generalCoefficients}
We now consider the final class of summary statistics: A weighted SFS with general coefficients. Recall that the BLUE from Section~\ref{BLUEsec} and the neutrality tests from Section~\ref{sec:neut} are examples of linear functions of the SFS with positive and negative coefficients. In this section we describe how to obtain the characteristic function of $\bc^\ast \bxi$, and how to invert the function to determine the cumulative distribution function (CDF) from numerical inversion. The inversion technique was used to determine the CDF for Tajima's~$D$ in Figure~\ref{fig:cdfapprox}, and is used below to determine the CDF for the BLUE from Section~\ref{BLUEsec}.  

If we define $z^\bc \eqdef (z^{c_1},z^{c_2},\dots,z^{c_{n-1}})^\ast$, we can obtain the PGF of $\bc^\ast \bxi$ from \eqref{eq:pgfSFS} as
\begin{align}
  G(z) \eqdef 
  \Exp [z^{\bc^\ast \bxi}] = 
  \Exp [z^{c_1 \xi_1} z^{c_2 \xi_2} \dots z^{c_{n-1} \xi_{n-1}}] =
  \be_1^\ast \left( -\lambda \Delta[\bA (z^\bc - \vect{e})] -\bT \right)^{-1} \vect{t}. \label{eq:firstversion}
\end{align}
where $\bA$ and $\bT$ are given in Theorem~\ref{thm:sfs} and $\vect{t}=-\mat{T}\vect{e}$.
We obtain the {\em characteristic function} of $\bc^\ast \bxi$ as
\begin{align}
  \phi(t)=G(\e^{\ih t}). 
\label{eq:phitildefun} 
\end{align}
We use numerical inversion techniques to obtain the cumulative distribution function~$F$. In \cite{waller1995} we find the inversion formula
$$
  F(x)= 
  \frac{1}{2} - 
  \int_{-\infty}^\infty \frac{\phi(t)}{2 \pi \ih t} \e^{- \ih t x} \dd t
$$
with the following approximation attributed to \cite{bohman1975numerical}:
\begin{align}
F_Z(z) = \frac{1}{2} + \frac{\eta z}{2 \pi} - \sum_{\substack{\nu = 1 - H \\ \nu \neq 0}}^{H-1}
\frac{\phi_Z(\eta \nu)}{2 \pi \ih \nu} \e^{- \ih \eta \nu z}. \label{eq:invnumchf}
\end{align}
The subscript $Z$ indicates that the random variable is assumed to be centered. The mean of $\bc^\ast \bxi$ is easily available, and therefore this assumption is not a limitation. 
Furthermore, $H$ and $\eta$ are parameters which together determine the accuracy and range of $F$ of the approximation. To facilitate computations
and achieve good accuracy we use a fast Fourier transform. We therefore rewrite \eqref{eq:invnumchf} in a form which is suitable for this purpose. First, we note that
$$
\sum_{\substack{\nu = 1 - H \\ \nu \neq 0}}^{H-1}
\frac{\phi_Z(\eta \nu)}{2 \pi \ih \nu} \e^{- \ih \eta \nu z}  =
\Re \left(\sum_{k = 1}^{H-1}
\frac{\phi_Z(\eta k)}{\pi \ih k} \e^{- \ih \eta k z}
\right)
$$
and setting  $z = 2 \pi h/(\eta H)$ in \eqref{eq:invnumchf} we find
\begin{align}
F\left(\frac{2 \pi h}{\eta H}\right) = \frac{1}{2} + \frac{h}{H}
+\Re \left(\sum_{k = 1}^{H-1}
\frac{\phi_Z(\eta k)}{\pi \ih k} \e^{- 2 \pi \ih k h/H }
\right),
\label{eq:ffteqn2}
\end{align}
from which we see that we can approximate $F$ in the interval $(-2 \pi/\eta,2 \pi/\eta)$ by calculating the fast 
Fourier transform $\hat{z}_h$ of the sequence
$$
z_h = \frac{\phi(\eta h)}{\pi \ih k}  \qquad h = 1-H,2-H,\dots,-2,-1,1,2\dots,H-2,H-1,
$$
and use formula \eqref{eq:ffteqn2} to obtain $F$ from $(\hat{z}_h)$. We have implemented the numerical inversion technique in the \texttt{phasty} package, and used it to determine the CDF for Tajima's~$D$ in Figure~\ref{fig:cdfapprox}.

We now discuss the BLUE from Section~\ref{BLUEsec}. Recall from~(\ref{BLUEestimator}) that the BLUE $\hat{\bc}$ is given by 
$\hat{\bc}=\bLambda^{-1}\bv/(\bv^{\ast}\;\bLambda^{-1}\bv)$ where $\bv$ is the vector with entries $v_i=1/i$ and $\mat{\Lambda} = \Var[\vect{\xi}]$ is the covariance matrix of~$\bxi$. We find the covariance matrix by combining equation~(\ref{eq:covmatdecomp}) and Theorem~\ref{thm:sfs}. In matrix notation
\begin{eqnarray*}
  \Var[\vect{\xi}]=
  \mat{\Lambda}=
  \mat{\Lambda}(\theta)=
  \frac{\theta^2}{4} \mat{\Sigma}+\frac{\theta}{2}\mat{\Delta}(\vect{\mu})=
  \frac{\theta^2}{4} \mat{\Sigma}+\theta \mat{\Delta}(\vect{\nu}).
\end{eqnarray*} 
Here entries in $\vect{\mu}$ are the means of the $i$-ton branch lengths (given by $2/i$) and entries in $\mat{\Sigma}$ are the covariances of the $i$-ton branch lengths which are calculated from Theorem~\ref{thm:sfs}, i.e. the fact that $i$-ton branch lengths are MPH-distributed. 

Note that for $\theta \rightarrow 0$ we get 
$\mat{\Lambda}(\theta) \approx \theta \mat{\Delta}(\vect{\nu})$ and 
$\hat{\vect{c}} \approx \vect{e}/(\vect{\nu}'\vect{e})=\vect{e}/a_1$, which is Watterson's estimator (\ref{WattersonEstimator}).  

\begin{figure}[h]
  \centering
  \includegraphics[scale=0.33]{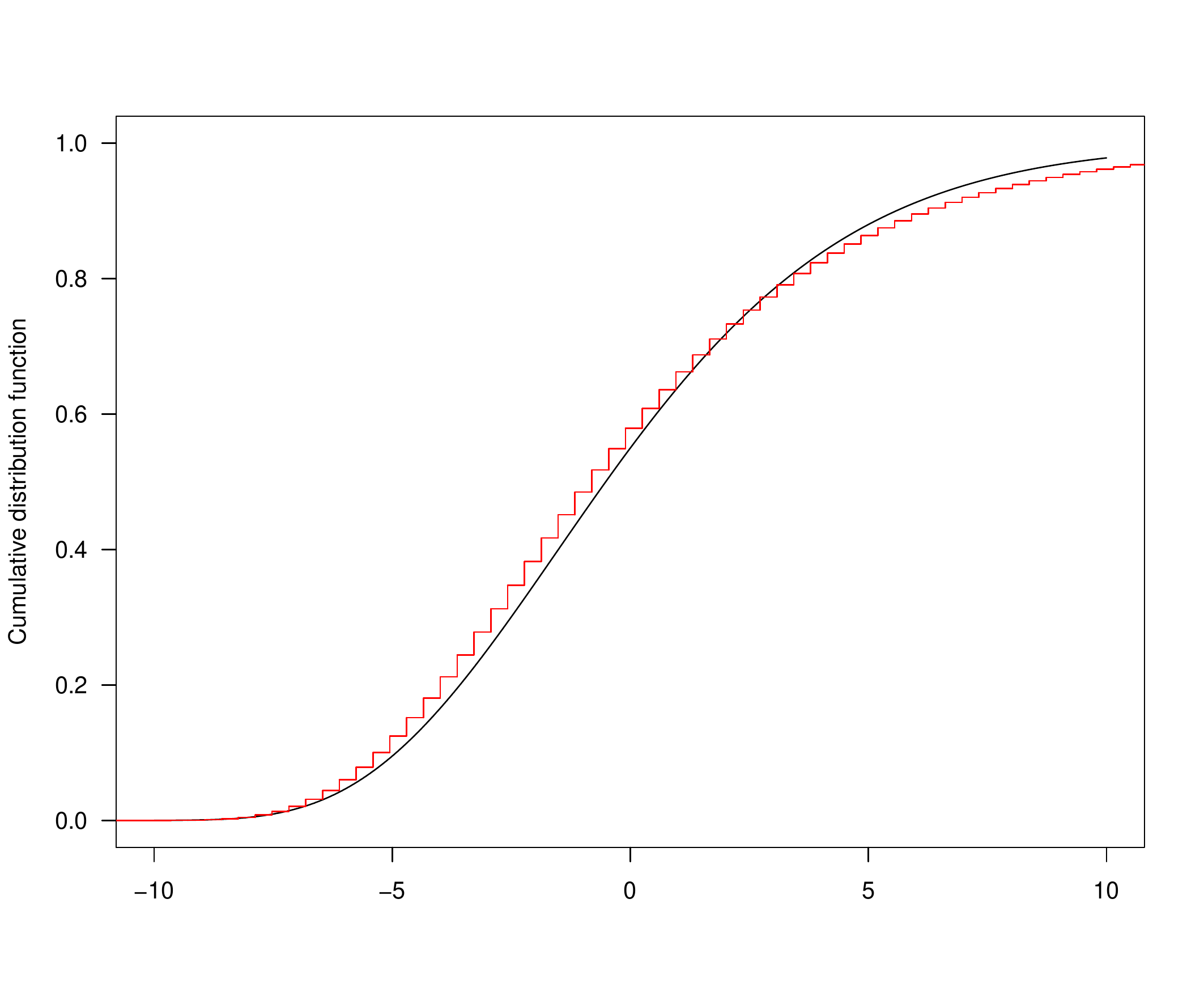}
  \includegraphics[scale=0.33]{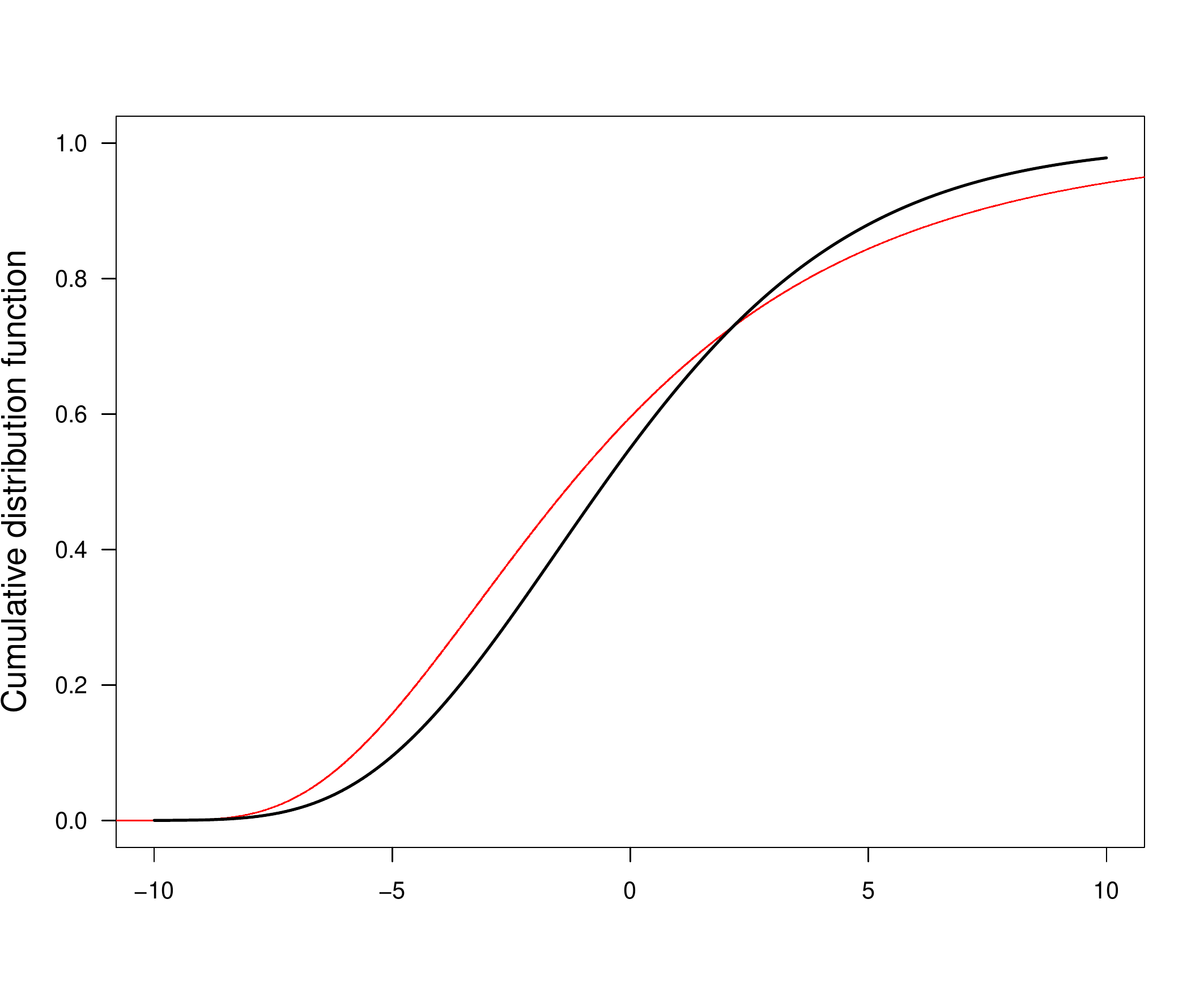}
  \caption{The CDF for the BLUE estimator (black) with the CDF for Watterson's estimator (left; in red) and the pairwise estimator (right; in red). The value of $\theta = 10$ and $n=10$.}
  \label{CDFFigure}
\end{figure}

In Figure~\ref{CDFFigure} we show the CDF for the BLUE estimator, Watterson's estimator and the pairwise difference estimator. All the estimators are unbiased (have mean $\theta$), and the mean is subtracted from the estimator. We can see from the plots that the BLUE estimator has the smallest variance and the CDF is rather smooth. Watterson's estimator also has a small variance, but the CDF is less smooth. Finally, the variance for the pairwise difference estimator is larger than for the other two estimators, but the CDF is again rather smooth.
\section{Software availability} 
Software implementation of phase--type methodology is available in the R package \phasty. The R package can be found at ({\verb+https://github.com/rivasiker/phasty+}). The figures and results in this paper are available as an accompanying vignette in the same repository.
\section{Conclusion and discussion} 
In this paper we have concentrated on the distribution of the joint site frequency spectrum for the standard coalescent with mutation (recall Theorem~\ref{thm:sfs}). Theorem~\ref{thm:sfs} is based on our general framework for sprinkling Poisson--distributed mutations on the MPH$^*$-distributed random variables (branch lengths) in Section~\ref{sect:PoissonOnMPH}, and can therefore be extended to more complex demographic scenarios as long as the ancestral process is homogeneous. The structured coalescent (see e.g. \cite{wakeley2008coalescent}, Section~5.2, \cite{Etheridge2012}, Chapter~6, and references therein), the coalescent with recombination (see e.g. \cite{wakeley2008coalescent}, Section~7.2 and references therein), and the multiple merger coalescent (see e.g. the recent review by \cite{BirknerBlath2019}) are examples of more general situations where our framework also applies. 

We have focused on the PGF, mean and (co)variance of the site frequency spectrum (SFS), but formulas are also available for e.g. the third--order (cross) moments of the SFS. The first--order moments for the branch lengths are given by~(\ref{eq:gen-mean}), the cross moments are given by~(\ref{eq:gen-cross-moment}), and higher--order moments are available from Theorem~8.1.5 in~\cite{bladt-nielsen-2017}. The generalization of the law of total expectation~(\ref{LawOfTotalExpectation}) and the law of total variance~(\ref{eq:covmatdecomp}) is the law of total cumulance (\cite{Brillinger1969}). The third--order moment of the Poisson distribution is analytically tractable which means that the third--order (cross) moments of the entries in the SFS are also analytically tractable. \cite{KlassmannFeretti2018} calculated the third moments of the site frequency spectrum by following and extending the theory outlined by \cite{FU1995172}. We advocate a more high--level approach based on manipulation of matrices.     

The distribution of Tajima's D is an example of a reward transformation with positive and negative rewards, and is often simulated. In our R package the function \texttt{rphtype} can be used to simulate the distribution of Tajima's D by simulating from the 2-dimensional discrete PH-type where e.g. the reward vector (1,-2,-1,3) is divided into (1,0,0,3) and (0,2,1,0), and then the two are subtracted in the end. 

The block matrix construction can result in very large matrices even for small sample sizes. However, the matrices are sparse and have much structure, and these two properties could be used to generally transfer the matrix manipulations to recursive formulae. This extension is left for future research.  
\section{Appendix} 
\subsection{General construction for positive integer-valued coefficients} \label{GenSectDPH}
In this section we provide the general construction of the DPH-representation of
a linear combination of the SFS with positive integer-valued coefficients, which was presented for several concrete cases in Section~\ref{integerCoefficients}. In particular for $6\hat{\theta}_{\pi}=3\xi_1+4\xi_2+3\xi_3$ in Figure~\ref{BlockDPHpairwiseFigure}, for $3\hat{\theta}_{\rm L}=\xi_1+2\xi_2+3\xi_3$ in eqn.~(\ref{ML}), and for $6\hat{\theta}_{\rm H}=\xi_1+4\xi_2+9\xi_3$ in eqn.~(\ref{MH}). 
Let
$\bc \in \mathbb{N}^{n-1}$ and consider
$$
\bc^\ast \bxi = c_1 \xi_1 + c_2 \xi_2 + \dots + c_{n-1} \xi_{n-1}
$$
Let $p \eqdef p(n)-1$ denote the size of the state space of the block counting process corresponding to sample size $n$ and let $\bA$ denote the $p \times (n-1)$ 
matrix, where the rows of $\bA$, which we will refer to as $\ba_i^\ast, i=1,\dots,p$, $\ba_i \in \mathbb{N}_0^{n-1}$ constitute an enumeration of the state-space of the block-counting process introduced in Section~\ref{01coefficients}. As an example, for $n=4$, we have
$$
\bA = \begin{Bmatrix}
4 & 0 & 0 \\
2 & 1 & 0 \\
1 & 0 & 1 \\
0 & 2 & 0 \\
\end{Bmatrix}.
$$
Let $\br = (r_1,\dots,r_p)$ denote the reward vector for the total reward in each state, i.e. $r_i = \sum_{j=1}^{n-1} c_{ij}$, and let 
$$
  \bM \eqdef \big(\bI_{p} - \frac{2}{\theta} \Delta (\br)^{-1} \bS\big)^{-1} 
  \eqdef \{p_{ij}\}.
$$
The transition matrix $\tilde{\bM}$ of the DPH-representation for the integer--valued SFS is given as a block matrix
$$
  \tilde{\bM} \eqdef \{ \tilde{\bM}_{ij} \} = 
  \begin{Bmatrix}
  \tilde{\bM}_{11} & \tilde{\bM}_{12} & \dots   & \tilde{\bM}_{1 m} \\
  \tilde{\bM}_{21} & \tilde{\bM}_{22} & \dots   & \tilde{\bM}_{2 m} \\
  \vdots           & \vdots            & \ddots & \vdots \\
  \tilde{\bM}_{m1} & \tilde{\bM}_{12} & \dots   & \tilde{\bM}_{m m} \\
  \end{Bmatrix} \, .
$$
The blocks below the diagonal i.e. $\tilde{\bM}_{ij}$ with $i>j$ are identically $0$.
 
Consider a diagonal-block $\tilde{\bM}_{ii}$. Such a block is a square matrix whose size is the maximal coefficient $a_j$ such that a jump in state $i$, can generate a 
$j$-ton., i.e., the maximal $a_j$ such that $c_{ij}$ is greater than zero:
$$
\mathfrak{m}\eqdef \max_{j = 1,\dots,n-1} \{c_j \indi{a_{ij}>0}\}.
$$
The matrix $\bM_{ii}$ has itself a block structure, namely
$$
  \tilde{\bM}_{ii}= 
  \begin{Bmatrix} 
    \bzero & \bI_{\mathfrak{m}-1} \\
    \tilde{\bbm}_{1} & \tilde{\bbm}_{2:m}
\end{Bmatrix}
$$
Let $\tilde{\bbm}$ denote the bottom row of $\tilde{\bM}_{ii}$.
Then $\tilde{\bbm} = p_{ii} \bbm$ where, informally, 
$\bbm$ gives the weights for the number of transitions (taken as sums of the entries from $\ba_i$) normalized so that the sum of the entries is $1$. Formally,
we first consider the un-normalized entries of $\bbm$.
Here the $(\mathfrak{m}-k)$th entry is the sum of the weights $a_{ij}$ for which $c_j = k$ that is
$$
  m_{\mathfrak{m}-k+1} = \sum_{j=1}^{n-1} a_{ij} \indi{c_j = k} \, ,
$$
where  $\bbm = (m_1,\dots,m_\mathfrak{m})$.
The entries of $\bbm$ are then normalized by their sum.

Finally consider the blocks above the diagonal i.e. $\bM_{ij}$ with $i<j$.
These blocks are identically zero, with the exception of the bottom row, which has the
form $p_{ij} \bbm$ where the $\bbm$ vector is the weight-vector calculated for diagonal-block $\tilde{\bM}_{jj}$.
\subsection{Derivatives of matrices} \label{sec:derivativesofC}
Let $\bU(z)$ and $\bV(z)$ denote two matrices of compatible orders.
The product rule extended to matrices says that
\begin{align}
\left(\bU(z) \bV(z)\right)' = \bU'(z) \bV(z) + \bU(z) \bV'(z). \label{eq:prodmatrule}
\end{align}
It follows that 
\begin{eqnarray*}
  0=\frac{\dd}{\dd z} \Big( \bU(z)) \bU(z)^{-1} \Big)
   =\Big( \frac{\dd}{\dd z} \bU(z) \Big) \bU(z)^{-1}+
    \bU(z) \frac{\dd}{\dd z} \Big( \bU(z)^{-1} \Big),
\end{eqnarray*}
and we get 
\begin{eqnarray*}
   \frac{\dd}{\dd z} \Big( \bU(z)^{-1} \Big)=
   -\bU(z)^{-1} \Big( \frac{\dd}{\dd z} \bU(z) \Big) \bU(z)^{-1}.
\end{eqnarray*}
In particular if $\bV$ is a constant we get
\begin{align}
  \frac{\dd}{\dd z} (\bV + \bU(z))^{-1} 
  =
  - \left(\bV + \bU(z)\right)^{-1}   \bU(z)' \left(\bV + \bU(z)\right)^{-1}. 
\label{DerivSumInverse}
\end{align}
\bibliographystyle{plainnat}
\bibliography{SegregatingSitesBib}
\end{document}